\documentclass[a4paper,12pt,leqno]{article}

\usepackage{latexsym}
\usepackage{bm,amsfonts,amsmath,amssymb} 
\usepackage[utf8]{inputenc}
\usepackage{abstract}
\usepackage[T1]{fontenc}
\usepackage{graphicx}
\usepackage{verbatim}
\usepackage{ulem}

\usepackage{epsfig}
\usepackage{color}

\usepackage[perpage]{footmisc}
\usepackage{fancybox}
\usepackage{dsfont}

\begin{document}
\newcommand{\beq}{\begin{equation}}
\newcommand{\eeq}{\end{equation}}

\newcommand{\R}{\mathbb{R}}
\newcommand{\C}{\mathbb{C}}
\newcommand{\N}{\mathbb{N}}
\newcommand{\Z}{\mathbb{Z}}
\newcommand{\Q}{\mathbb{Q}}
\newcommand{\mj}{\mathcal}
\newcommand{\pa}{\partial}

\newcommand{\rg}{\mathrm{rg}}
\newcommand{\su}{\mathrm{supp}}
\newcommand{\la}{\lambda}
\newcommand{\fcv}{\rightharpoonup}
\newcommand{\bg}{\begin}
\newcommand{\ds}{\displaystyle}
\newcommand{\Om}{\Omega}
\newcommand{\eps}{\epsilon}
\newcommand{\Sp}{\mathbb{S}}
\newcommand{\inj}{\hookrightarrow}
\newcommand{\n}{\textbf{n}}
\newcommand{\bv}{\textbf{b}}
\newcommand{\h}{\textbf{h}}
\newcommand{\A}{\textbf{A}}
\newcommand{\F}{\textbf{F}}
\newcommand{\B}{\textbf{B}}
\newcommand{\dr}{\partial}
\newcommand{\vv}{\textbf{v}}
\newcommand{\uu}{\textbf{u}}
\newcommand{\tr}{\mathrm{Tr}}
\newcommand{\conj}{\overline}
\newcommand{\intn}{\int_{\Om}\!\!\!\!\!\!\!-}
\newcommand{\dive}{\mathrm{div}}
\newcommand{\bbeta}{\bm{\beta}}
\newcommand{\vect}{\overrightarrow}

\newtheorem{lem}{Lemma}[section]
\newtheorem{theo}[lem]{Theorem}
\newtheorem{prop}[lem]{Proposition}
\newtheorem{sch}[lem]{Scholie}
\newtheorem{cor}[lem]{Corollary}

\newenvironment{rem}
{\noindent\addtocounter{lem}{1}
{\textbf{Remark \thelem.}}\\\noindent\rm}
{\bg{flushright}\tiny $\blacksquare$\end{flushright}}

\newenvironment{preuve}
{\noindent{\textbf{Proof.}}\\\rm\noindent}
{\bg{flushright}\tiny $\blacksquare$\end{flushright}}

\def\got#1{{\bm{\mathfrak{#1}}}}

\renewcommand{\theequation}{\thesection.\arabic{equation}}
\renewcommand{\thefigure}{\thesection.\arabic{equation}}

\title{On the semi-classical 3D Neumann Laplacian with variable magnetic field}
\author{Nicolas Raymond}
\maketitle{}
\bg{abstract}
In this paper we are interested in the semi-classical estimates of the spectrum of the Neumann Laplacian in dimension 3. This work aims to present a complementary case to the one presented in the paper of Helffer and Morame in the case of constant magnetic field. More precisely, in the case when the magnetic field is variable and under the most generic condition for which boundary localizations can be observed, we prove a three terms upper bound for the lowest eigenvalue and establish some semi-classical behaviour of the spectrum.
\end{abstract}

\section{Introduction and main results}
Let $\Om$ be an open bounded subset of $\R^3$ with smooth boundary. For a vector potential $\A\in\mj{C}^{\infty}(\conj{\Om},\R^3)$ and $h>0$, we introduce 
$$\bbeta=\nabla\times\A$$ 
and the quadratic form defined for all $\psi\in H^1(\Om,\C)$ by :
$$q_\A^h(\psi)=\int_\Om |(ih\nabla+\A)\psi|^2 dx.$$
One of our interests will be the lowest eigenvalue denoted by $\la_1^h(\A)$ of the associated self-adjoint operator i.e. the Neumann realization of $(ih\nabla+\A)^2$ on $\Om$ denoted by $P_\A^h$.
The question of analyzing the behaviour of $\la_1^h(\A)$ as $h$ tends to $0$ appears in the theory of superconductivity when studying the third critical field of the Ginzburg-Landau functional (see \cite{FouHel2}) and also in the theory of phase transition of liquid crystals (see \cite{HelPan}). In this last topics, it turns out that the case when the magnetic field admits constant strength really plays a particular role. In this paper, we will treat the case when both direction and strength of the magnetic field may vary, what can happen in the superconductivity case. In fact, until now, only the case with constant magnetic field was studied in details.
Before analyzing the general problem, we need to recall some properties of a simplified model on the half space which were first established in \cite{Lupan2} and then in \cite{HelMo, FouHel2}.
\paragraph{Model in $\R^3_+$}~\\
Let us consider the case of the constant magnetic field in 
$$\R_+^3=\{(r,s,t)\in\R^3 : t>0\}.$$  
The angle between the magnetic field $\bbeta$ and the plane $t=0$ is denoted by $\theta_1$ and so, up to a rotation, we can assume (in the case when the norm of $\bbeta$ is $1$) : 
$$\bbeta=(0,\cos\theta,\sin\theta),$$
with $$\theta=\theta_1\in[0,\frac \pi 2].$$
An associated vector potential is given by :
$$\A=(t\cos\theta-s\sin\theta,0,0)$$
and we consider the Neumann realization on $\R^3_+$ of
\beq\label{Hrtheta}
\mathcal H(\theta)=(D_r+V_\theta)^2+D_s^2+D_t^2,
\eeq
where 
\beq\label{V}
V_\theta=t\cos\theta-s\sin\theta.
\eeq
The bottom of the spectrum of this operator is denoted by :
\bg{equation}\label{sigmatheta}
\inf\sigma(\mathcal H(\theta))=\sigma(\theta).
\end{equation}
Atlthough we work in 3D, we will have to consider the Neumann realization on $\R^2_+$ denoted by $H(\theta)$ with domain :
$$D(H(\theta))=\{\psi\in L^2(\R^2_+) : H(\theta)\psi\in L^2(\R^2_+) \mbox{ and } D_t\psi=0 \mbox{ on } t=0\}$$
and defined by :
\beq\label{Htheta}
H(\theta)=D_t^2+D_s^2+V_\theta^2.
\eeq
It has been proved in \cite{Lupan2} (see \cite{FouHel2} and the references therein), that :
\bg{enumerate}
\item $\inf\sigma(H(\theta))=\sigma(\theta)$ for $\theta\in\left]0,\frac{\pi}{2}\right[$,
\item $\sigma$ is analytic on $\left]0,\frac{\pi}{2}\right[$, strictly increasing on $\left[0,\frac{\pi}{2}\right]$ and satisfies : ${0<\sigma(0)<1}$ and $\sigma\left(\frac{\pi}{2}\right)=1$.
\item For $\theta\in\left]0,\frac{\pi}{2}\right[$~:
\beq\label{sigmaess}
\inf \sigma_{ess}( H(\theta)) =1\,.
\eeq
\item $\sigma(\theta)$ is a simple eigenvalue of $H(\theta)$ associated to some positive and $L^2$-normalized function $u_\theta$ (see also \cite{ReSi}).
\end{enumerate}
In particular, the minimal energy of $\mj{H}(\theta)$ is obtained when the magnetic field is tangent ($\theta=0$).
Using a rescaling, the bottom of the spectrum when the field admits $\|\bbeta\|$ for module is given by $\|\bbeta\|\sigma(\theta).$\\
For $x\in\dr\Om$, we let :
\bg{equation}\label{betachapeau}
\hat{\bbeta}(x)=\sigma(\theta(x))\|\bbeta(x)\|,
\end{equation}
where $\theta(x)$ is defined by :
$$\|\bbeta(x)\|\sin\theta(x)=\bbeta\cdot\nu(x)$$
with $\nu(x)$ the inward pointing normal at $x$.

\paragraph{Lu-Pan's result : main term of the asymptotics}~\\
We recall the asymptotics which was first established by Lu-Pan (cf. \cite{Lupan2,FouHel2}) :
\bg{theo}\label{vrough3}
The following asymptotic expansion holds :
$$\la_1^h(\A)=\min(\inf_{\Om}\|\bbeta(x)\|,\inf_{\dr\Om}\hat{\bbeta}(x))h+O(h^{5/4}).$$
\end{theo}
\bg{rem}
\vspace*{-0.5cm}
\bg{enumerate}
\item In fact, until now, the proof of the upper bound was just sketched (see \cite[Theorem 5.1]{Lupan2}). We will see that, if some generic conditions are satisfied, the remainder for the upper bound is better than $O(h^{5/4})$.
\item The Stokes formula implies that :
$$\int_{\dr\Om}\bbeta\cdot\nu\, d\sigma=\int_\Om \nabla\cdot\bbeta\, dx=0,$$
and so there exists a point where the magnetic field is tangent to the boundary. Consequently, when the magnetic field has constant strength, the first term in the asymptotics is $\Theta_0\|\bbeta(x_0)\|h$.
\item The case when the magnetic field is constant has already been studied in details under generic assumptions in \cite{HelMo2} where a two terms asymptotics is proved with a second term of order $h^{4/3}$.
\item When the magnetic field just admits a constant strength, one or two term asymptotics are established in the papers \cite{HelPan, HelPan2, Ray}. 
\item Finally, let us mention that this kind of asymptotics has already appeared in dimension 2 in the constant case (see \cite{BeSt, BPT,Lupan, PFS, HelMo3, FouHel3}) and that we have obtained a two terms expansion in \cite{Ray2} when the magnetic field is not uniform.
\end{enumerate}
\end{rem}
\paragraph{Main assumptions and ideas}~\\
Let us state our hypotheses on the magnetic field in this paper. 
Our estimates are only interesting when the following condition is satisfied (cf. Theorem (\ref{vrough3})) :
\beq\label{boundary}
\inf_{\dr\Om}\hat{\bbeta}<\inf_{\Om}\|\bbeta\|.
\eeq
This condition is also called the surface superconductivity condition (cf. \cite{Lupan2, FouHel2}). It implies a localization of the groundstates on the boundary when $h\to 0$. Hence, on the boundary, we wish to deal with some generic case in the same way as we have done in 2D (see \cite{Ray2}) ; so we will assume first that :
\beq\label{minimum0}
\hat{\bbeta} \mbox{ admits a minimum at } x_0
\eeq 
and that at this point the field is neither tangent, neither perpendicular to the boundary, i.e :
\beq\label{angle}
0<\theta(x_0)=\theta_1<\frac{\pi}{2}.
\eeq
Our point is to obtain the semi-classical behaviour of eigenvalues in the bottom of the spectrum of $P_\A^h$. In order to do that we will construct an operator whom spectrum is close to the one of $P_\A^h$. From this, it will follow a quasimode construction associated to the approximated operator and simultaneously we will get, in some sense, the candidate to a three terms asymptotics (in powers of $h$) of the lowest eigenvalues. To define the different terms in the asymptotics, each one corresponding to a step of the quasimode construction, we need to introduce a few invariants mixing the geometry of the domain, the properties of the magnetic field and spectral quantities attached to $H(\theta)$.
We first define the direct orthonormal basis $(\vect{l_0}, \vect{\tau_0}, \vect{\nu_0})$ attached to the point $x_0$~:
\bg{enumerate}
\item $\vect{\nu_0}$ is the inward pointing normal at $x_0$
\item $\ds{\vect{\tau_0}=\frac{\bbeta(x_0)-(\bbeta(x_0)\cdot\vect{\nu_0})\vect{\nu_0}}{\|\bbeta(x_0)-(\bbeta(x_0)\cdot\vect{\nu_0})\vect{\nu_0}\|}}$
\item $\vect{l_0}=\vect{\tau_0}\times\vect{\nu_0}$.
\end{enumerate}
Let us now present our main results.
\paragraph{Second term of the asymptotics}~\\
Our first invariant depends on the the second fundamental form at $x_0$ denoted by $K$ :
$$K_{11}=K(\vect{l_0},\vect{l_0}), K_{22}=K(\vect{\tau_0},\vect{\tau_0}), K_{12}=K(\vect{l_0},\vect{\tau_0}).$$
It also depends on the first derivatives of $\bbeta$ at $x_0$ :
$$\beta=\nabla<\frac{\bbeta}{\|\bbeta(x_0)\|},\vect{\tau_0}>\cdot\vect{\tau_0}$$
$$\tilde{\delta}=\nabla<\frac{\bbeta}{\|\bbeta(x_0)\|},\vect{l_0}>\cdot\vect{l_0},$$
$$\eta=\nabla<\frac{\bbeta}{\|\bbeta(x_0)\|},\vect{\tau_0}>\cdot\vect{\nu_0}-K_{11}\cos(\theta),$$
with $\theta=\theta_1$.
At $x_0$ we define the quantity $C^{\bbeta,K}(x_0)$ by :
\bg{align}\label{CBK}
2\beta< V_\theta(st-\frac{T}{2}s^2)u_\theta, u_\theta>+\eta <t^2 V_\theta  u_\theta,u_\theta>+\frac{2\tilde{\delta}}{\sin\theta}\int_{t>0} t|D_s u_\theta|^2+<H_K u_\theta,u_\theta>,\nonumber\\
\end{align}
where $V_\theta$ is defined in (\ref{V}) and
\beq\label{defT}
T=T(\theta)=-\frac{C(\theta)}{S(\theta)},
\eeq
with
\beq\label{defCS}
C=C(\theta)=\cos\theta\sigma(\theta)-\sin\theta\sigma'(\theta)\,,
S=S(\theta)=\sin\theta\sigma(\theta)+\cos\theta\sigma'(\theta)\,
\eeq
and
\beq\label{HK}
H_K=(K_{11}+K_{22}) \dr_t+2tK_{11}V_\theta^2+2tK_{22}D_s^2+2tK_{12}(V_\theta D_s+D_s V_\theta).
\eeq
\bg{rem}
\bg{enumerate}
\vspace*{-0.5cm}
\item We can notice that $S(\theta)>0$. Indeed, we know (see \cite{FouHel2}) that $\sigma'(\theta)\geq 0$, $\sigma(\theta)>0$ and we have assumed that $\theta\neq 0$.
\item An integration by parts provides that $$<t(V_\theta D_s+D_s V_\theta)u_\theta,u_\theta>=0$$ and so $C^{\bbeta,K}(x_0)$ is real.
\item We will see that the construction of $C^{\bbeta,K}(x_0)$ as part of the second term of the asymptotics uses that $x_0$ is a critical point of $\hat{\bbeta}$.
\end{enumerate}
\end{rem}
We can now state a theorem dealing with the second term of the asymptotics~:
\bg{theo}\label{majgen0}
Under Assumptions (\ref{minimum0}) and (\ref{angle}), there exists $D>0$ and $h_0>0$ such that, for $0<h\leq h_0$, there exists at least one eigenvalue $\mu$ of $P_\A^h$ such that :
$$|\mu-(\hat{\beta}(x_0)h+C^{\bbeta,K}(x_0)\|\bbeta(x_0)\|^{1/2}h^{3/2})|\leq D h^{2},$$
where $C^{\bbeta,K}(x_0)$ is defined in (\ref{CBK}).
\end{theo}
Then, to continue the asymptotic expansion, we assume the following generic condition :
\beq\label{minimum}
\hat{\bbeta} \mbox{ admits a non-degenerate minimum at } x_0.
\eeq 
\paragraph{Third term of the asymptotics}~\\
We denote by $\got{S}_{\beta}$ the Hessian matrix at $x_0$, namely :
$$\got{S}_{\bbeta}(r,s)=\frac{1}{2}\dr_r^2\hat{\bbeta}(x_0)r^2+\frac{1}{2}\dr^2_{rs}\hat{\bbeta}(x_0)(sr+rs)+\frac{1}{2}\dr_s^2\hat{\bbeta}(x_0)s^2$$
We can now introduce a fundamental operator linked with the behaviour of the third term of the asymptotics :
\beq\label{osc}
\tilde{\got{S}}_{\bbeta}=\got{S}_{\bbeta}(D_\tau,\frac{\tau}{\sin\theta}).
\eeq
By the non-degeneracy Assumption (\ref{minimum}), this operator is an harmonic oscillator whose spectrum is explicitly known. We can now state the main result of this paper :
\bg{theo}\label{majgen}
Under Assumptions (\ref{minimum}) and (\ref{angle}), there exists $d\in\R$ such that for all $n\in\N^*$, there exists $D_n>0$ and $h_n>0$ such that, for $0<h\leq h_n$, there exists at least one eigenvalue $\mu_n$ of $P_\A^h$ such that :
$$|\mu_n-(\hat{\beta}(x_0)h+C^{\bbeta,K}(x_0)\|\bbeta(x_0)\|^{1/2}h^{3/2}+(\gamma_n(\tilde{\got{S}}_{\bbeta})+d)h^2)|\leq D_n h^{5/2},$$
where $\gamma_n(\tilde{\got{S}}_{\bbeta})$ is the $n$-th eigenvalue of $\tilde{\got{S}_{\bbeta}}$.
\end{theo}
This theorem admits the following easy corollary :
\bg{cor}
Under Assumptions (\ref{minimum}) and (\ref{angle}), there exists $d\in\R$, $D_1>0$ and $h_1>0$ such that, for $0<h\leq h_1$ :
$$\la_1^h(\A)\leq \hat{\beta}(x_0)h+C^{\bbeta,K}(x_0)\|\bbeta(x_0)\|^{1/2}h^{3/2}+(\gamma_1(\tilde{\got{S}}_{\bbeta})+d)h^2+D_1 h^{5/2}.$$
\end{cor}
An interesting point would be to get the corresponding lower bound.
\paragraph{Organization of the paper}~\\
In Section 2, we study some spectral properties of $H(\theta)$. In Section 3, we consider a change of coordinates and see how the magnetic field and the metrics are transformed. In Section 4, we explain how we get, using Taylor approximations, a simplified operator. Finally, in Section 5, we use the simplified model to construct a quasimode and prove Theorem \ref{majgen0} and \ref{majgen}.

\section{The family of operators $H(\theta)$}
This section is devoted to the study of a family of operators $H(\theta)$ which plays a fundamental role. Let us explain why we are reduced to this family of operators in $\R^2_+$. Considering the operator (\ref{Hrtheta}) and performing a Fourier transform in the variable $r$, we are led to the $\tau$-dependent family of Neumann realizations :
$$H(\theta,\tau)=(\tau+V_\theta)^2+D_s^2+D_t^2.$$ 
Then, under Assumption (\ref{angle}) and making a translation in the variable $s$, $H(\theta,\tau)$ is unitarily equivalent to $H(\theta,0)=H(\theta)$.\\
In the next subsection, we study the exponential decay properties of the solutions of equations in the form :
$$h(\theta)v=f,$$
where : 
\beq\label{htheta}
h(\theta)=H(\theta)-\sigma(\theta).
\eeq
This kind of equations will indeed appear many times in the final quasimode construction.
\subsection{Exponential decay}
We begin to notice that, in view of (\ref{sigmaess}) and using Persson's Theorem (cf. \cite{Agmon,FouHel2}), when $0<\theta<\frac{\pi}{2}$ and for any $0<\alpha<\sqrt{1-\sigma(\theta)}$, we get :
$$e^{\alpha\sqrt{s^2+t^2}}u_\theta \in H^1(\R^2_+).$$
Let us denote 
$$L^2_{exp}(\R^2_+)=\{f\in L^2(\R_+^2) : \exists\alpha>0,\quad e^{\alpha(t+<s>)}f\in L^2(\R^2_+)\}$$
and 
$$H^1_{exp}(\R^2_+)=\{f\in L^2(\R_+^2) : \exists\alpha>0,\quad e^{\alpha(t+<s>)}f\in H^1(\R^2_+)\},$$
where $<s>=(s^2+1)^{1/2}$.\\
Thus, we have :
\beq\label{exputheta}
u_\theta\in H_{exp}^1(\R_+^2).
\eeq
Let us now state the main proposition of this subsection.
\bg{prop}\label{dec}
Let $f\in L^2_{exp}(\R^2_+) $ such that $<f\,,\,u_\theta>=0$.
Then, if $v$ denotes a solution of 
\beq\label{eq}
h(\theta)v=f, 
\eeq
then $$v\in H_{exp}^1(\R_+^2).$$
\end{prop}
\bg{preuve}
We begin to prove the control in the variable $s$. Let us introduce a smooth cutoff function $\chi$ satisfying 
$$\chi(s)=1 \mbox{ if } |s|\leq 1 \mbox{ and } \chi(s)=0 \mbox{ if } |s|\geq 2.$$
For $n\in\N^*$, we consider the (bounded and inversible) multiplication operator by $e^{-\eps\chi_n(s)<s>}$ where $\chi_n(s)=\chi(n^{-1}s).$ Let us verify that it preserves $D(H(\theta))$.\\
For $\psi\in D(H(\theta))$, we have :
$$H(\theta)(e^{-\eps\chi_n(s)<s>}\psi)=e^{-\eps\chi_n(s)<s>}H(\theta)\psi-\eps\chi'_n(s)<s>D_s \psi-\eps\chi_n(s)\frac{s}{<s>}D_s \psi.$$
As $D_s \psi\in L^2(\R_+^2)$, we deduce that $H(\theta)(e^{-\eps\chi_n(s)<s>}\psi)\in L^2(\R_+^2).$ Moreover, $e^{-\eps\chi_n(s)<s>}\psi$ satisfies the Neumann condition and so :
$$e^{-\eps\chi_n(s)<s>}\psi\in D(H(\theta)).$$
We now use a Grushin method and we introduce $\got{H}(\theta)$ defined on $D(H(\theta))\times\C$ by :
$$\got{H}(\theta)=
\left[\bg{array}{cc}
h(\theta)&u_\theta\\
<\cdot\, ,u_\theta>&0
\end{array}\right].$$
It is obvious that this operator is inversible.
(\ref{eq}) is equivalent to :
$$\got{H}(\theta)\left[\bg{array}{c}
v\\
0
\end{array}\right]=\left[\bg{array}{c}
f\\
0
\end{array}\right],$$
if we take for $v$ the unique solution orthogonal to $u_\theta$ (all the other solutions will satisfy the decay properties because $u_\theta$ does).
Letting $\tilde{v}=e^{\eps\chi_n(s)<s>}v$ and $\tilde{f}=e^{\eps\chi_n(s)<s>}f$, we can rewrite :
$$\got{H}(\theta)^{\eps,n}\left[\bg{array}{c}
\tilde{v}\\
0
\end{array}\right]=\left[\bg{array}{c}
\tilde{f}\\
0
\end{array}\right],$$
where
$$\got{H}(\theta)^{\eps,n}=\left[\bg{array}{cc}
e^{\eps\chi_n(s)<s>}&0\\
0&1
\end{array}\right]\got{H}(\theta)\left[\bg{array}{cc}
e^{-\eps\chi_n(s)<s>}&0\\
0&1
\end{array}\right].$$
We can notice that $D(\got{H}(\theta)^{\eps,n})=D(\got{H}(\theta)).$
An easy computation provides :
$$\got{H}(\theta)^{\eps,n}=\left[\bg{array}{cc}
e^{\eps\chi_n(s)<s>}h(\theta)e^{-\eps\chi_n(s)<s>}&e^{\eps\chi_n(s)<s>}u_\theta\\
<\cdot\,,\, e^{-\eps\chi_n(s)<s>} u_\theta>&0
\end{array}\right].$$
We can write, using the exponential decay of $u_\theta$ (see (\ref{exputheta})), that :
$$\got{H}(\theta)^{\eps,n}=\got{H}(\theta)+\eps F_{n},$$
where $F_n$ is bounded relatively to $\got{H}(\theta)$ uniformly with respect to $n$.
The conclusion is then standard ; for $\eps$ small enough $\got{H}(\theta)^{\eps,n}$ is inversible and there exists $C(\eps)>0$ such that for all $n\in\N^*$ :
$$\|(\got{H}(\theta)^{\eps,n})^{-1}\|\leq C(\eps).$$
This leads to the inequality :
$$\|\tilde{v}\|_{L^2(\R_+^2)}\leq C(\eps)\|\tilde{f}\|_{L^2(\R_+^2)}.$$
Thus, we deduce :
$$\|e^{\eps\chi_n(s)<s>}v\|_{L^2(\R_+^2)}\leq C(\eps)\|e^{\eps\chi_n(s)<s>}f\|_{L^2(\R_+^2)}.$$
The dominated convergence Theorem shows that the r.h.s converges when $n\to+\infty$ to 
$\|e^{\eps<s>}f\|_{L^2(\R_+^2)}.$
So, we infer that $e^{\eps\chi_n(s)<s>}v$ weakly converges (after subsequence extraction) in $L^2(\R^2_+)$ to some function $h$. The convergence in the sense of distributions and again the dominated convergence theorem proves that : ${e^{\eps<s>}v=h\in L^2(\R^2_+).}$
Consequently, the decay with respect to $s$ is proved.\\
We now explain the decay with respect to the variable $t$. We proceed exactly in the same way, but we have to deal with the Neumann condition. For $n\in\N^*$, we consider a smooth cutoff function $\eta_n$ such that :
$$
\eta_n(t)=\left\{
\bg{array}{cc}
0& \mbox{ if } 0\leq t\leq\frac{1}{2}\\
1& \mbox{ if } 1\leq t\leq n\\
0& \mbox{ if } t\geq 2n
\end{array}\right..
$$
We observe that the (bounded and inversible) multiplication operator by $e^{\eps\eta_n(t)t}$ preserves $D(H(\theta))$ ; then, it remains to use the same analyis as previously by using this time the decay of $u_\theta$ with respect to $t$ and the proposition is proved after having noticed (by integration by parts) that :
$$q_\theta(e^{\alpha/2(t+<s>)}v)<+\infty,$$
where $q_\theta$ is the  quadratic form associated to $H(\theta)$.
\end{preuve}
Let us denote :
$$H^\infty_{exp}=\{f\in L^2(\R_+^2) : \forall (\ell,k)\in\N^4\quad D_s^\ell D^k_t f\in  L^2_{exp}(\R^2_+)\}.$$
It is clear that $H^\infty_{exp}\subset\mj{S}(\conj{\R_+\times\R})$.
Proposition \ref{dec} permits to prove the following proposition :
\bg{prop}
The groundstate $u_\theta$ of $H(\theta)$ belongs to $H^\infty_{exp}$.
\end{prop}
\bg{preuve}
We first recall that $u_\theta\in H^1_{exp}(\R^2_+)$.
Then, let us take the derivative with respect to $s$ :
$$h(\theta)\dr_s u_\theta=2\sin\theta V_\theta u_\theta\in L^2_{exp}(\R^2_+).$$ 
Proposition \ref{dec} provides $D_s u_\theta\in H^1_{exp}(\R^2_+)$.
A recursion gives that, for all $m\in\N$ : 
$$D^m_s u_\theta\in H^1_{exp}(\R^2_+).$$
Then, we come back to the equation :
$$D_t^2 u_\theta=(-D_s^2-V_\theta^2+\sigma(\theta))u_\theta\in L^2_{exp}(\R^2_+).$$
Taking the derivative with respect to $s$, we find :
$$D_t^2 D_s^m u_\theta\in L^2_{exp}(\R^2_+)$$
and differentiating with respect to $t$, we deduce :
$$D_t^3 D_s^m u_\theta\in L^2_{exp}(\R^2_+).$$
Finally, a recursion provides, for all integers $\ell, k$ : 
$$D_s^\ell D^k_t u_\theta \in L_{exp}^2(\R_+^2).$$
\end{preuve}
We can now state the other important corollary :
\bg{cor}\label{dec2}
Let $g\in H^\infty_{exp}$ and $f\in D(H(\theta))$ such that :
$$h(\theta)f=g.$$
Then, $f\in H^\infty_{exp}$.
\end{cor}
\bg{preuve}
The proof is essentially the same as the one of the previous corollary.
\end{preuve}

\subsection{A few properties of $u_{\theta}$}
We begin to establish some properties of $u_\theta$ related to the tangential coordinate $s$.
\subsubsection{A momentum formula}
The definition of $u_{\theta}$ provides :
$$H(\theta)u_{\theta}=\sigma(\theta)u_{\theta}\,.$$
We differentiate with respect to $s$ this identity and obtain the formula :
\bg{equation}\label{ds}
h(\theta)\dr_s
u_{\theta}=2\sin\theta\, V_\theta u_{\theta}.
\end{equation}
Therefore, taking the scalar product with $u_{\theta}$, we deduce the lemma :
\bg{lem}\label{moment0}~\\
For all $\theta\in]0,\frac{\pi}{2}[$, we have :
$$\int_{t>0}(t\cos\theta-s\sin\theta)|u_{\theta}(s,t)|^2  ds dt=0\,.$$
\end{lem}
We again take the derivative in (\ref{ds}) with respect to $s$ to find :
\bg{equation}\label{ds2}
h(\theta)\dr^2_s
u_{\theta}=2\sin\theta (2V_\theta \dr_s u_\theta-\sin\theta u_{\theta})=2\sin\theta(\dr_s V_\theta+V_\theta \dr_s)u_\theta.
\end{equation}
If we take the derivative two more times, we get :
$$h(\theta)\dr^4_s u_\theta=12\sin^2\theta \dr_s^2 u_\theta+8\sin\theta V_\theta \dr_s^3 u_\theta.$$
We deduce the lemma :
\bg{lem}\label{ds4}
We have :
$$\int V_\theta \dr_s^3 u_\theta u_\theta dsdt=\frac{3}{2}\sin\theta\int \dr^2_s u_\theta u_\theta dsdt.$$
\end{lem}
\subsubsection{Another formulas}
We will meet the question of finding $g$ s.t :
\bg{equation}\label{g}
h(\theta)g=-t\dr_s u_\theta.
\end{equation}
We let :
\beq\label{Ls}
L_s=-\dr_s^2+\sin^2\theta\, s^2.
\eeq
An easy computation gives the following commutator formula :
$$L_s H(\theta)=H(\theta)L_s+4\cos\theta\sin\theta t\dr_s.$$
We infer :
$$h(\theta)(L_s u_\theta)=-4\cos\theta\sin\theta t\dr_s u_\theta.$$
Hence, we can define a solution of (\ref{g}) by :
\beq\label{f_0}
f_0=(2\sin (2\theta))^{-1}(\sin^2\theta s^2 u_\theta-\dr_s^2 u_\theta).
\eeq
Finally, we have :
$$L_sh(\theta)f_0=h(\theta)(L_sf_0)+4\cos\theta\sin\theta\, t\dr_sf_0.$$
We take the scalar product to get :
$$<L_sh(\theta)f_0\,,\,u_\theta>=<4\cos\theta\sin\theta t\dr_sf_0\,,\,u_\theta>.$$
But, we have :
\bg{align*}
&<L_sh(\theta)f_0\,,\,u_\theta>=-<L_s(t\dr_s u_\theta)\,,\,u_\theta>=-\sin^2\theta\int ts^2\,\dr_s u_\theta\, u_\theta\, ds dt\\
&=\sin^2\theta\int ts u_\theta^2 ds dt.
\end{align*}
So, we deduce the following lemma :
\bg{lem}\label{if0}
We have the identity :
$$<4\cos\theta t\dr_sf_0,u_\theta>=\sin\theta\int ts u_\theta^2 ds dt.$$
\end{lem}
Finally, differentiating twice the equation (\ref{g}) satisfied by $f_0$, we obtain :
\bg{lem}\label{Vf0}
We have, on one hand :
$$h(\theta)\dr_sf_0=-t\dr^2_s u_\theta+2\sin\theta V_\theta f_0,$$
$$<-t\dr^2_s u_\theta+2\sin\theta V_\theta f_0,u_\theta>=0$$
and on the other hand :
$$h(\theta)\dr_s^2 f_0=-t\dr_s^3 u_\theta+2\sin\theta(-\sin\theta f_0+2 V_\theta \dr_s f_0),$$
$$<-t\dr_s^3 u_\theta+2\sin\theta(-\sin\theta f_0+2 V_\theta \dr_s f_0),u_\theta>=0.$$
\end{lem}

\subsubsection{Properties of $H(\theta,\rho)$}
The main ingredient in the following analysis is the so-called Feynman-Hellmann formula.
We introduce a parameter $\rho$ which will permit to play with a scale invariance.
We define :
$$H(\theta,\rho)=\frac{1}{\rho}(D_s^2+D_t^2)+\rho V_\theta^2$$
and consider :
\bg{equation}\label{vp}
H(\theta,\rho)u_{\theta,\rho}=\sigma(\theta)u_{\theta,\rho},
\end{equation}
where 
$$u_{\theta,\rho}(s,t)=u_{\theta}(\rho^{-1/2} s,\rho^{-1/2} t).$$
We can notice that $H(\theta,1)=H(\theta)$.
\subsubsection{Feynman-Hellmann with respect to $\rho$}
We take the derivative with respect to $\rho$ :
$$(H(\theta,\rho)-\sigma(\theta))\dr_\rho u=-\dr_\rho H u,$$
and
$$\dr_\rho H=-\rho^{-2}(D_s^2+D_t^2)+\rho^{-1/2}(\tau+\sqrt{\rho}V_\theta)V_\theta.$$
We use (\ref{vp}) to obtain :
\bg{equation}\label{drhou}
(H(\theta,\rho)-\sigma)\dr_\rho u=\frac{\sigma u_{\theta,\rho}}{\rho}-2V_\theta^2 
u_{\theta,\rho}.
\end{equation}
Multiplying this last equation by $u_{\theta,\rho}$ and integrating, one recovers the so-called Virial Theorem :
\bg{equation}\label{virial}
\int_{t>0}\left(|D_s u_\theta|^2+|D_t u_\theta|^2 \right) ds dt=\int_{t>0}|V_\theta
u_\theta|^2 ds dt=\frac{\sigma(\theta)}{2}\,.
\end{equation}
\bg{rem}
If we only perform a rescaling in the variable $s$, we find the identity :
\beq\label{virial2}
\int |D_s u_\theta|^2 ds dt=-\sin\theta\int sV_\theta u_\theta^2 ds dt.
\eeq
\end{rem}
\subsubsection{Feynman-Hellmann with respect to $\theta$}
We take the derivative of (\ref{vp}) with respect to $\theta$  :
$$(H(\theta,\rho)-\sigma(\theta))\dr_\theta u=\sigma'(\theta)u-\dr_\theta H u,$$
and
$$\dr_\theta H=2\rho V_\theta\dr_\theta V_\theta.$$
We find :
\bg{equation}\label{dthetau}
(H-\sigma)\dr_\theta u=\sigma'(\theta)u_{\theta}-2\rho V_\theta\dr_\theta V_\theta u_{\theta}.
\end{equation}
Taking $\rho=1$ and multiplying by $u_{\theta}$, we get :
\bg{equation} \label{FHtheta}
2\int_{t>0}V_\theta\frac{\dr V_\theta}{\dr\theta}|u_\theta|^2 ds
dt=\sigma'(\theta)\,,
\end{equation}
which also reads :
\bg{equation}
\sigma'(\theta):= 
 - 2 \int_{t>0} (\cos \theta \, t -\sin \theta \, s) (\sin \theta \, t + \cos
 \theta\, s) |u_\theta (s,t)|^2 ds
dt\,.
\end{equation}
\bg{rem}
Combining (\ref{FHtheta}) and (\ref{virial}), it follows :
\beq\label{tVsV}
\int t V_\theta u_\theta^2 dsdt=\frac{C(\theta)}{2}\quad \mbox{and}\quad \int s V_\theta u_\theta^2 dsdt=-\frac{S(\theta)}{2}.
\eeq
\end{rem}
\subsubsection{Consequences}
\paragraph{A first identity}~\\
We get the following lemma :
\bg{lem}\label{I1}
Defining
\bg{equation}\label{uta1}
I_1(\theta) =\int_{t>0}(t-sT(\theta))V_{\theta}|u_{\theta}|^2 ds dt\,,
\end{equation}
where $T(\theta)$ is defined in (\ref{defT}), we have : $I_1(\theta)=0.$
\end{lem}
\bg{preuve}
One multiplies (\ref{drhou}) by $\sigma'$ and (\ref{dthetau}) by $\frac{\sigma}{\rho}$ to obtain :
\beq\label{wthetarho}
(H(\theta,\rho)-\sigma(\theta))w_{\theta,\rho}=-2\sigma'V_\theta^2 u_{\theta,\rho}+2\sigma V_\theta\dr_\theta V_\theta u_{\theta,\rho},
\eeq
where
\bg{equation}
w_{\theta,\rho}=\sigma' \dr_\rho u_{\theta,\rho}-\frac{1}{\rho}\sigma \dr_\theta u_{\theta,\rho}.
\end{equation}
We take $\rho=1$, multiply by $u_{\theta}$, integrate and Lemma \ref{I1} is proved.
\end{preuve}
We will denote :
\bg{equation}\label{wtau}
w_0=\sigma' \dr_\rho u_{\theta,1}-\sigma \dr_\theta u_{\theta,1}.
\end{equation}

\paragraph{A second identity}~\\
We now apply the operator 
\beq\label{j}
j(\theta,\rho)=\sigma'\dr_\rho-\frac{\sigma}{\rho}\dr_\theta
\eeq
to (\ref{wthetarho}), take $\rho=1$ and we obtain :
\bg{align*}
&j(\theta,\rho)(H(\theta,\rho))w_{\theta,\rho}+(H(\theta,\rho)-\sigma)j(\theta,\rho)w_{\theta,\rho}\\
&=(-2\sigma' V_\theta^2+2\sigma V_\theta \dr_\theta V_\theta)w_{\theta,\rho}+j(\theta,\rho)(-2\sigma' V_\theta^2+2\sigma V_\theta \dr_\theta V_\theta) u_{\theta,\rho}.
\end{align*}
Then, on the one hand, we get :
\beq\label{jH}
j(\theta,\rho)(H(\theta,\rho))=-\sigma'(D_s^2+D_t^2)+\sigma' V_\theta^2-2\sigma V_\theta \dr_\theta V_\theta,
\eeq
and one the other hand :
\bg{align*}
&j(\theta,\rho)(-2\sigma' V_\theta^2+2\sigma V_\theta \dr_\theta V_\theta)=-\sigma\dr_\theta(-2\sigma' V_\theta^2+2\sigma V_\theta \dr_\theta V_\theta)\\
&=2\sigma''\sigma V_\theta^2+2\sigma\sigma'V_\theta\dr_\theta V_\theta-2\sigma^2(\dr_\theta V_\theta)^2-2\sigma^2 V_\theta \dr_\theta^2 V_\theta.
\end{align*}
Taking the scalar product with $u_\theta$, and using (\ref{FHtheta}) and (\ref{virial}), we deduce the lemma :
\bg{lem}\label{w0i}
We have the identity :
$$\int_{t>0}(2\sigma' V_\theta^2-2\sigma V_\theta \dr_\theta V_\theta)w_0 u_\theta ds dt+\int_{t>0}(\sigma\dr_\theta V_\theta-\sigma'V_\theta)^2 u_\theta^2 ds dt=\frac{\sigma^2\sigma''+\sigma^3}{2}.$$
\end{lem}
\paragraph{A third identity}~\\
If we take the derivative of (\ref{wthetarho}) two times with respect to $s$ , we get :
\bg{align*}
h(\theta)\dr_s w_0=2\sin\theta V_\theta w_0-2V_\theta u_\theta+2(Cs+St)\sin\theta u_\theta-2(Cs+St)V_\theta \dr_s u_\theta
\end{align*}
and then :
\bg{align*}
&h(\theta)\dr^2_s w_0=4\sin\theta V_\theta\dr_s w_0-2\sin^2\theta w_0-2(Cs+St)V_\theta \dr_s^2 u_\theta\\ 
&+4C\sin\theta u_\theta -4CV_\theta\dr_s u_\theta+4\sin\theta(Cs+St)\dr_s u_\theta
\end{align*}
and we deduce the lemma :
\bg{lem}\label{ds2w0}
We have on one hand :
\bg{align*}
&2\sin\theta<V_\theta w_0,u_\theta>+2\sin\theta<(Cs+St)u_\theta,u_\theta>-2<(Cs+St)V_\theta\dr_s u_\theta,u_\theta>=0
\end{align*}
and on the other hand :
\bg{align*}
&<4\sin\theta V_\theta\dr_s w_0-2\sin^2\theta w_0-2(Cs+St)V_\theta \dr_s^2 u_\theta,u_\theta>\\
&=-\sin\theta<(2C+4(Cs+St)\dr_s) u_\theta,u_\theta>=0.
\end{align*}
\end{lem}

\paragraph{Two more identities satisfied by $f_0$ and $w_0$}~\\
If we rescale the equation (\ref{g}) satisfied by $f_0$, we get :
$$(H(\theta,\rho)-\sigma(\theta))f_{0,\rho}=-t\dr_s u_{\theta,\rho},$$
where 
$$f_{0,\rho}(s,t)=f_0(\rho^{-1/2}s,\rho^{-1/2}t).$$
We now apply the operator $j$, take $\rho=1$ and use the identity (\ref{jH}) to get :
$$h(\theta)j(f_{0,\rho})+\sigma' t\dr_s u_\theta+(2\sigma'V_\theta^2-2\sigma V_\theta \dr_\theta V_\theta)f_0=-t\dr_s w_0.$$
Taking the scalar product with $u_\theta$, we find :
$$\int (2\sigma'V_\theta^2-2\sigma V_\theta \dr_\theta V_\theta)f_0 u_\theta ds dt+\int t\dr_s w_0 u_\theta ds dt=0.$$
Moreover, noticing that 
$$0=j\left(\int t\dr_su_{\theta,\rho}^2 ds dt\right)=2\int tu_{\theta,\rho}w_{\theta,\rho} dsdt,$$
we get the lemma :
\bg{lem}\label{crossed}
We have the following cancellations :
$$\int (2\sigma'V_\theta^2-2\sigma V_\theta \dr_\theta V_\theta)f_0 u_\theta ds dt=0$$
and
$$\int t\dr_s w_0 u_\theta ds dt=0$$
\end{lem} 
\bg{rem}\label{dec3}
It follows from Corollary \ref{dec2} that all the solutions of equations of the form $h(\theta)f=g$
that we have met in this section belong to $H^\infty_{exp}.$
\end{rem}
\section{A choice of coordinates near the boundary}
This section deals with a system of coordinates near the boundary. Indeed, it will be useful to be reduced to a simplified model on an half space.
\subsection{A general choice of coordinates}
Let us assume that $0\in\dr\Om$. In some neighborhood $V$ of $0$, we take the coordinates $(y_1,y_2)$ on $\dr\Om$ (via a map $\mj{C}^3$ denoted $\phi$ and which will be defined precisely in the following). We denote $\nu(\phi^{-1}(y_1,y_2))$ the inward pointing normal  at the point $\phi^{-1}(y_1,y_2)$ and we define local coordinates in $V$ :
$$\Phi(y_1,y_2,y_3)=\phi^{-1}(y_1,y_2)+y_3 \nu(\phi^{-1}(y_1,y_2)).$$
More precisely, for a point $x\in V$, $\phi^{-1}(y_1,y_2)$ is the projection of $x$ on $\dr\Om\cap V$ and $y_3=t=d(x,\dr\Om)$.
\paragraph{Transformation of the magnetic field}~\\
We now want to determine the new vector potential and magnetic field in these new coordinates.
Let us introduce the 1-form $\omega$ :
$$\omega=A_1 dx_1+A_2 dx_2+A_3 dx_3.$$
We have :
$$\omega=\tilde{A}_1 dy_1+\tilde{A}_2 dy_2+\tilde{A}_3 dy_3.$$
We can express $dx_i$ as a function of $(dy_j)$ and we get :
$$\tilde{\A}=D_y\Phi^{-1}(\A(\Phi^{-1}(y))).$$
Then, in order to find the new field, we write :
$$d\omega=(\nabla\times\A)_1 dx_2\wedge dx_3+(\nabla\times\A)_2 dx_1\wedge dx_3+(\nabla\times\A)_3 dx_1\wedge dx_2.$$
The comatrix formula provides :
\beq\label{nb}
\tilde{\bbeta}=\det(D\Phi)^{-1}((D\Phi))^t\bbeta,
\eeq
where
$$\tilde{\bbeta}=\nabla_y\times\tilde{\A}.$$ 
\paragraph{Metrics in the new coordinates}~\\
The euclidean metrics is
$$g_0=dx_1^2+dx_2^2+dx_3^2.$$ 
In the new coordinates $(y_1,y_2,y_3)$, we have :
$$g_0=\sum_{k,j}g_{ij}dy_k\otimes dy_j,$$
where $(g_{kj})$ is the matrix $(D\Phi^{-1})^t (D\Phi^{-1})$.
We denote $(g^{kj})$ the inverse matrix of $(g_{kj})$.\\
If the support of $u$ is sufficiently concentrated near $x_0$, we have :
\beq
q_{\A}^h(u)=\int_{\Om}|(ih\nabla+\A)u|^2dx=\int_{t>0}|(ih\nabla_y+\tilde{\A})\tilde{u}|_{D\Phi{}^t(D\Phi)}^2 |\det D\Phi^{-1}| dy,
\eeq
where
$$|(ih\nabla_y+\tilde{\A})\tilde{u}|_{D\Phi{}^t(D\Phi)}^2=\sum_{k,j}g^{kj}(ih\nabla_{y_k}+\tilde{A}_k)\tilde{u}\,\conj{(ih\nabla_{y_j}+\tilde{A}_j)\tilde{u}}.$$
We will denote by $\tilde{P}_\A^h$ the associated Neumann realization on $L^2(|\det D\Phi^{-1}| dy)$, namely :
\beq\label{mq}
\tilde{P}_\A^h=|\det D\Phi|\sum_{k,j}(ih\nabla_j+\tilde{A}_j)|\det D\Phi^{-1}| g^{kj}(ih\nabla_k+\tilde{A}_k).
\eeq
\subsection{Coordinates $(r,s,t)$}\label{rst}
We now introduce normal coordinates near $x_0$. $(\vec{\tau_0},\vec{l_0})$ is an orthonormal basis of the tangent plane at $x_0$ ; we denote by $(r,s)$ the corresponding coordinates. Then, it is standard, using the exponential map near $x_0$, that $(r,s)$ define a local parametrization of the boundary and they are called normal coordinates (cf. \cite{Lf}). So, there exists an open set $S$ of $\R^2$ and a diffeomorphism $\phi$ such that :
$$\phi : \mj{W}_{x_0}\to S, \phi(x)=(r,s).$$
Moreover, if $G$ denotes the first fundamental form of $\dr\Om$, we have :
$$G=I_3+G_1(r,s)+O(r^3+s^3),$$
where $G_1$ is a quadratic form in $r$ and $s$.\\
Letting $t(x)=d(x,\dr\Om)$, we define the system of coordinates on the boundary~: $(y_1,y_2,y_3)=(r,s,t)$.
The metrics $g_0$ can be expressed as :
$$g_0=dt\otimes dt+G-2tK+t^2 L,$$
where $K, L$ are respectively the second and third fundamental forms on $\dr\Om$.
\paragraph{Taylor expansion of the metrics}~\\
We let :
$$K_0=\left(\bg{array}{cc}
K_{11}&K_{12}\\
K_{12}&K_{22}\\
\end{array}\right)$$
We write the followimg Taylor expansion :
$$K=K_0+K_1(r,s)+O(r^2+s^2),$$
where $K_1$ is linear in $r$ and $s$.\\ 
In general, for a $2\times2$ matrix $U$, we will denote $U^0$ the matrix :
$$U^0=\left(\bg{array}{ccc}
U_{11}&U_{12}&0\\
U_{12}&U_{22}&0\\
0&0&0\\
\end{array}\right).$$
Thus, we can write :
\bg{align*}
&g_0=I_3-2tK^0+G^0_1(r,s)-2tK_1^0+t^2L^0+O(r^3+s^3)\\
&=I_3-2tK^0+G^0_1(r,s)+R(r,s,t)+O(r^3+s^3+t^3)\\
&=g_{app}+O(r^3+s^3+t^3),
\end{align*}
where $R$ is homogeneous of degree $2$ and of partial degree $1$ with respect to $r$ and $s$.\\ 
\bg{rem}\label{R}
In the following, we take as convention to always denote by $R$ the homogeneous polynoms of degree $2$ and of partial degree $1$ with respect to $r$ and $s$.
\end{rem}
We deduce :
$$D\Phi^{-1}=I_3-tK^0+\frac{G_1^0}{2}+R(r,s,t)+O(r^3+s^3+t^3)$$
and
$$D\Phi=I_3+tK^0-\frac{G_1^0}{2}+R(r,s,t)+O(r^3+s^3+t^3).$$
It follows :
$$|g|^{1/2}=1-tK^M_0+\tr(\frac{G_1^0}{2})+R(r,s,t)+O(r^3+s^3+t^3)=m_{app}+O(r^3+s^3+t^3),$$
where $K_0^M=K_{11}+K_{22}$ is the mean curvature and :
$$m_{app}=|g_{app}|^{1/2}.$$
The dual metrics satisfies :
\beq\label{mdap}
g^0=I_3+2tK^0-G^0_1(r,s)+R(r,s,t)+O(r^3+s^3+t^3)=g^0_{app}+O(r^3+s^3+t^3).
\eeq
\bg{rem}\label{notation}
In the following, it will be convenient to denote by $G_{ij}^{r^2}$ (resp. $G_{ij}^{s^2}$, $G_{ij}^{rs}$) the coefficient of $r^2$ in the coefficient with index $(i,j)$ of $G_1$ (resp. the coefficient of $s^2$, the coefficient of $rs$). 
\end{rem}
\section{An approximated operator on $\R^3_+$}
In this section, we use the local coordinates introduced in the previous section to construct a model operator near $x_0$. The one term asymptotics was obtained by considering the model with constant field in $\R^3_+$ ; in the following, we keep more terms in the approximation of $P_{\A}^h$ to get a three terms asymptotics.
\subsection{The model}
In this subsection, we approximate first the metrics and then the magnetic field in order to get a model operator.
\paragraph{Approximation of the metrics}~\\
With (\ref{mq}) and (\ref{mdap}), we are reduced to consider the operator on $L^2(m_{app}drdsdt)$, with Dirichlet condition on $t=t_0$, $|r|=r_0$, $|s|=s_0$ (for $t_0$, $r_0$, $s_0$ positive and small enough) and Neumann condition on $t=0$ :
\bg{align*}
\mj{H}^{app}=&h^2(m_{app})^{-1}D_t m_{app} D_t\\
&+(1+2tK_{11}-G_{11}-R_{11})(hD_r+\tilde{A_r})^2+(1+2tK_{22}-G_{22}-R_{22})(hD_s+\tilde{A_s})^2\\
&+(2K_{12}t-G_{12}-R_{12})(hD_r+\tilde{A_r})(hD_s+\tilde{A_s})\\
&+(2K_{12}t-G_{12}-R_{12})(hD_s+\tilde{A_s})(hD_r+\tilde{A_r})\\
&+h^2\tilde{R}_1(r,s,t)D_r+h^2\tilde{R}_2(r,s,t)D_s,
\end{align*}
where the $\tilde{R}_i(r,s,t)$ are linear functions in $r$, $s$ and $t$ coming from a commutator between the metrics and the derivatives $D_r$ and $D_s$.\\
This first approximation of the operator $\tilde{P}_\A^h$ satisfies, for $\psi$ sufficiently concentrated near $x_0$ (an explicit choice will be made in the following) :
\bg{align}\label{rfqm}
&\|\tilde{P}_\A^h\psi-\mj{H}^{app}\psi\|\leq Ch^2\sum_{k,j}\|(|r|^3+|s|^3+t^3)|D_k D_j\psi|\|\\
\nonumber&+Ch\|(|r|^3+|s|^3+t^3)|\psi|\|+Ch\sum_{j}\|(r^4+s^4+t^4)|D_j\psi|\|+C\|(|r|^5+|s|^5+t^5)|\psi|\|.
\end{align}
Omitting the terms of order $3$, we get :
\bg{align}\label{Hmod}
&\mj{H}^{MOD}=h^2 n_{app}D_t m_{app}D_t\\
\nonumber&+(1+2tK_{11}-G_{11}-R_{11})(hD_r+\tilde{A_r})^2+(1+2tK_{22}-G_{22}-R_{22})(hD_s+\tilde{A_s})^2\\
\nonumber&+(2K_{12}t-G_{12}-R_{12})(hD_r+\tilde{A_r})(hD_s+\tilde{A_s})\\
\nonumber&+(2K_{12}t-G_{12}-R_{12})(hD_s+\tilde{A_s})(hD_r+\tilde{A_r})\\
\nonumber&+h^2\tilde{R}_1(r,s,t)D_r+h^2\tilde{R}_2(r,s,t)D_s.
\end{align}
with 
$$n_{app}=1+tK^M_0-\tr(\frac{G_1^0}{2})+R.$$
\paragraph{Taylor approximation of the magnetic field}~\\
The Taylor formula applied to $\bbeta$ at $x_0$ gives (at the order $1$):
\bg{align*}
&\bbeta=\bbeta_0+r\dr_r\bbeta(x_0)+s\dr_s\bbeta(x_0)+t\dr_t\bbeta(x_0)\\
&=\bbeta^{(1)}+O(r^2+s^2+t^2),
\end{align*}
and with (\ref{nb}) we are led to :
\bg{align*}
&\tilde{\bbeta}=\bbeta_0+t(-K_0^M\bbeta_0+K^0\bbeta_0+\dr_t\bbeta(x_0))+r\dr_r\bbeta(x_0)+s\dr_s\bbeta(x_0)+O(r^2+s^2+t^2).\\
&=\tilde{\bbeta}^{(1)}+O(r^2+s^2+t^2).
\end{align*}
We can notice that, on $t=0$, we have $\tilde{\bbeta}^{(1)}=\bbeta^{(1)}.$\\
Let us write :
\bg{align*}
&{\beta_r}^{(1)}=-\delta_0 r-\eps_0 s-\xi_0 t,\\
&{\beta_s}^{(1)}=\cos\theta_1+\alpha_0 r+\beta_0 s+\eta_0 t,\\
&{\beta_t}^{(1)}=\sin\theta_1+\gamma_0 s+\zeta_0 r+\mu_0 t,\\
\end{align*}
and
\bg{align*}
&\tilde{\beta_r}^{(1)}=-\delta r-\eps s-\xi t,\\
&\tilde{\beta_s}^{(1)}=\cos\theta_1+\alpha r+\beta s+\eta t,\\
&\tilde{\beta_t}^{(1)}=\sin\theta_1+\gamma s+\zeta r+\mu t.\\
\end{align*}
Then, we have the relations :
\bg{align}\label{an1}
&\delta_0=\delta, \eps_0=\eps, \alpha_0=\alpha, \beta_0=\beta, \gamma_0=\gamma, \zeta_0=\zeta,\\
&\xi=\xi_0-K_{12}\cos\theta, \eta=\eta_0-K_{11}\cos\theta, \mu=\mu_0-K_0^M\sin\theta.
\end{align}
We now write the formula at the order $2$ for $\beta$ and $\tilde{\beta}$, but only when $t=0$~; indeed, this will be enough in the following. So, the approximation at the order $2$ of $\beta$ and $\tilde{\beta}$ on $t=0$ can be written as :
\bg{align*}
&{\beta_r}^{(2)}=-\delta r-\eps s-\conj{C_0}rs-\conj{F_0}s^2-\conj{D_0}r^2,\\
&{\beta_s}^{(2)}=\cos\theta_1+\alpha r+\beta s+C_0rs+F_0s^2+D_0r^2,\\
&{\beta_t}^{(2)}=\sin\theta_1+\gamma s+\zeta r-2B_0rs-3H_0s^2-E_0r^2.\\
\end{align*}
and
\bg{align*}
&\tilde{\beta_r}^{(2)}=-\delta r-\eps s-\conj{C}rs-\conj{F}s^2-\conj{D}r^2,\\
&\tilde{\beta_s}^{(2)}=\cos\theta_1+\alpha r+\beta s+Crs+Fs^2+Dr^2,\\
&\tilde{\beta_t}^{(2)}=\sin\theta_1+\gamma s+\zeta r-2Brs-3Hs^2-Er^2.\\
\end{align*}
Then, we have (cf. \ref{nb}), on $t=0$ :
$$\tilde{\beta}=\left(1+\tr(\frac{G_1}{2})\right)\left(I_3-\frac{G_1^0}{2}\right)\beta$$
and this provides a relation between the capital letters of $\tilde{\beta}$ and $\beta$.
Identifying the terms in $s^2$, we get (cf. Remark \ref{notation}):
\bg{align}\label{an}
&\conj{F}=\conj{F_0}+\frac{1}{2}\cos\theta G_{12}^{s^2},\\
\nonumber&F=F_0+\frac{1}{2}\cos\theta G_{11}^{s^2},\\
\nonumber&H=H_0-\frac{1}{6}\sin\theta (G_{11}^{s^2}+G_{22}^{s^2}).
\end{align}
Then, we obtain for the term in $r^2$ :
\bg{align}\label{an2}
&\conj{D}=\conj{D_0}+\frac{1}{2}\cos\theta G_{12}^{r^2},\\
\nonumber&D=D_0+\frac{1}{2}\cos\theta G_{11}^{r^2},\\
\nonumber&E=E_0-\frac{1}{2}\sin\theta (G_{11}^{r^2}+G_{22}^{r^2}).
\end{align}
and for the term in $rs$ :
\bg{align}\label{an3}
&\conj{C}=\conj{C_0}+\frac{1}{2}\cos\theta G_{12}^{rs},\\
\nonumber&C=C_0+\frac{1}{2}\cos\theta G_{11}^{rs},\\
\nonumber&B=B_0-\frac{1}{4}\sin\theta (G_{11}^{rs}+G_{22}^{rs}).
\end{align}
Then, we choose a gauge such that :
$$\tilde{A}_t=0$$
and such that the Taylor approximation of $\tilde{\A}$ at the order $3$ denoted by $\tilde{A}^{(3)}$ is defined by :
\bg{align*}
&\tilde{A_r}^{(3)}=V_\theta+P_{r,2}+P_{r,3}\\
&\tilde{A_s}^{(3)}=P_{s,2}+P_{s,3}\\
&\tilde{A_t}^{(3)}=0,
\end{align*}
where
\bg{align}\label{polys}
\nonumber&P_{r,2}=(\alpha t-\zeta s)r+\beta st-\frac{\gamma}{2}s^2+\frac{\eta}{2}t^2\\
&P_{r,3}=(At^2+Bs^2+Cst)r+(Dt+Es)r^2+(Fts^2+G t^2s+Hs^3+It^3)\\
\nonumber&P_{s,2}=\delta rt+\eps st+\frac{\xi}{2}t^2\\
\nonumber&P_{s,3}=\conj{A}t^2r+\conj{C}str+\conj{D}tr^2+\conj{F}ts^2+\conj{G}t^2s+\conj{I}t^3.
\end{align}
The model operator on $L^2(m_{app}dr ds dt)$ is : 
\bg{align}\label{Hmodele}
&\mj{H}^{\mj{M}}=h^2 n_{app}D_t m_{app}D_t\\
\nonumber&+(1+2tK_{11}-G_{11}-R_{11})(hD_r+\tilde{A_r}^{(3)})^2+(1+2tK_{22}-G_{22}-R_{22})(hD_s+\tilde{A_s}^{(3)})^2\\
\nonumber&+(2K_{12}t-G_{12}-R_{12})(hD_r+\tilde{A_r}^{(3)})(hD_s+\tilde{A_s}^{(3)})\\
\nonumber&+(2K_{12}t-G_{12}-R_{12})(hD_s+\tilde{A_s}^{(3)})(hD_r+\tilde{A_r}^{(3)})\\
\nonumber&+h^2\tilde{R}_1(r,s,t)D_r+h^2\tilde{R}_2(r,s,t)D_s
\end{align}
and it satisfies :
\bg{align}\label{rfqf}
\|\mj{H}^{app}\psi-\mj{H}^{\mj{M}}\psi\|\leq Ch\sum_{j}\|(r^4+s^4+t^4)D_j\psi\|+C\|(|r|^5+|s|^5+t^5)|\psi|\|.
\end{align}
In conclusion, we can write :
\bg{align}\label{approximation}
&\|\tilde{P}_\A^h\psi-\mj{H}^{\mj{M}}\psi\|\leq Ch^2\sum_{k,j}\|(|r|^3+|s|^3+t^3)|D_k D_j\psi|\|\\
\nonumber&+Ch\|(|r|^3+|s|^3+t^3)|\psi|\|+Ch\sum_{j}\|(r^4+s^4+t^4)|D_j\psi|\|+C\|(|r|^5+|s|^5+t^5)|\psi|\|.
\end{align}

\subsection{Non-degenerate minimum of $\hat{\bbeta}$}
In this subsection, we express the condition of non-degenerate minimum of $\hat{\bbeta}$. Firstly, we write that $x_0$ is a critical point of $\hat{\bbeta}$ and secondly we give the expression of the Hessian matrix of $\hat{\bbeta}$. Moreover, we will assume that $\|\bbeta(x_0)\|=1$.
\subsubsection{Critical point of $\hat{\bbeta}$}
By definition of $\theta$, we have :
\beq\label{deftheta}
\|\bbeta\|\sin\theta=\beta_t.
\eeq
Let us compute the partial derivatives of $\theta$.
We have :
\bg{align*}
&\dr_r\beta_t=\frac{\dr_r\bbeta\cdot\bbeta}{\|\bbeta\|}\sin\theta+\|\bbeta\|\cos\theta\,\dr_r\theta\\
&\dr_s\beta_t=\frac{\dr_s\bbeta\cdot\bbeta}{\|\bbeta\|}\sin\theta+\|\bbeta\|\cos\theta\,\dr_s\theta\\
\end{align*}
In $r=s=t=0$, we get :
\bg{align*}
&\zeta=(\alpha\cos\theta_1+\zeta\sin\theta_1)\sin\theta_1+\cos\theta_1\dr_r\theta\\
&\gamma=(\beta\cos\theta_1+\gamma\sin\theta_1)\sin\theta_1+\cos\theta_1\dr_s\theta\\
\end{align*}
and so :
\bg{align*}
&\dr_r\theta=\zeta\cos\theta_1-\alpha\sin\theta_1\\
&\dr_s\theta=\gamma\cos\theta_1-\beta\sin\theta_1.\\
\end{align*}
Then we differentiate $\hat{\bbeta}$ with respect to $r$ and $s$ :
\bg{align*}
&\dr_r\hat{\bbeta}=\frac{\dr_r\bbeta\cdot\bbeta}{\|\bbeta\|}\sigma(\theta)+\|\bbeta\|\sigma'(\theta)\dr_r\theta\\
&\dr_s\hat{\bbeta}=\frac{\dr_s\bbeta\cdot\bbeta}{\|\bbeta\|}\sigma(\theta)+\|\bbeta\|\sigma'(\theta)\dr_s\theta\\
\end{align*}
At $r=s=t=0$, we have : $\dr_r\hat{\bbeta}=0$ and $\dr_s\hat{\bbeta}=0$. This leads to :
\bg{align*}
&(\alpha\cos\theta_1+\zeta\sin\theta_1)\sigma(\theta_1)+\sigma'(\theta_1)\dr_r\theta=0\\
&(\beta\cos\theta_1+\gamma\sin\theta_1)\sigma(\theta_1)+\sigma'(\theta_1)\dr_s\theta=0\\
\end{align*}
Thus, we find :
\bg{align*}
&C(\theta_1)\alpha+S(\theta_1)\zeta=0,\\
&C(\theta_1)\beta+S(\theta_1)\gamma=0,\\
\end{align*}
where $C$ and $S$ are defined in (\ref{defCS}).
This can be rewritten as : 
\beq\label{zetagamma}
\zeta=T\alpha\mbox{  and }\gamma=T\beta\,,
\eeq
with $T$ defined in (\ref{defT}).
\subsubsection{Hessian matrix of $\hat{\bbeta}$}
A computation associated with the relations (\ref{zetagamma}) gives, for the first derivatives at $x_0$:
\bg{align*}
&\dr_r\|\bbeta\|=(\alpha\cos\theta+\zeta\sin\theta)=\alpha\frac{\sigma'}{S},\\
&\dr_s\|\bbeta\|=(\beta\cos\theta+\gamma\sin\theta)=\beta\frac{\sigma'}{S},
\end{align*}
and for the second derivatives ar $x_0$:
\bg{align*}
&\dr_r^2\|\bbeta\|=\delta^2+\frac{\sigma^2}{S^2}\alpha^2+2D_0\cos\theta-2E_0\sin\theta,\\
&\dr_s^2\|\bbeta\|=\eps^2+\frac{\sigma^2}{S^2}\beta^2+2F_0\cos\theta-6H_0\sin\theta\\
&\dr_{rs}^2\|\bbeta\|=\delta\eps+\alpha\beta\frac{\sigma^2}{S^2}+C_0\cos\theta-2B_0\sin\theta.
\end{align*}
So, we get the following Taylor expansion :
\bg{align*}
&\|\bbeta\|=1+\alpha\frac{\sigma'}{S}r+\beta\frac{\sigma'}{S}s+\frac{r^2}{2}(\delta^2+\frac{\sigma^2}{S^2}\alpha^2+2D_0\cos\theta-2E_0\sin\theta)\\
&+\frac{s^2}{2}(\eps^2+\frac{\sigma^2}{S^2}\beta^2+2F_0\cos\theta-6H_0\sin\theta)+rs(\delta\eps+\frac{\sigma^2}{S^2}\alpha\beta+C_0\cos\theta-2B_0\sin\theta).
\end{align*}
We now want a Taylor expansion of $\theta$. We have :
\bg{align*}
&\sin\theta=\sin\theta_1+(\cos\theta_1 \dr_r\theta+\sin\theta_1 \dr_s\theta)+r^2\left(\frac{\cos\theta_1}{2}\dr^2_r\theta-\frac{\sin\theta_1}{2}(\dr_r\theta)^2\right)\\
&+s^2\left(\frac{\cos\theta_1}{2}\dr^2_s\theta-\frac{\sin\theta_1}{2}(\dr_s\theta)^2\right)+rs\left(\cos\theta_1\dr_{rs}\theta-\sin\theta_1\dr_r\theta\dr_s\theta\right).
\end{align*}
After computations using (\ref{deftheta}), we get : 
\bg{align*}
&\frac{\dr_r^2\theta}{2}=\alpha^2\frac{\sigma\sigma'}{S^2}-D_0\sin\theta-E_0\cos\theta-\frac{\sin\theta}{2\cos\theta}\delta^2,\\
&\frac{\dr_s^2\theta}{2}=\beta^2\frac{\sigma\sigma'}{S^2}-F_0\sin\theta-3H_0\cos\theta-\frac{\sin\theta}{2\cos\theta}\eps^2,\\
&\dr^2_{rs}\theta=-2\alpha\beta\frac{\sigma\sigma'}{S^2}-C_0\sin\theta-2B_0\cos\theta-\frac{\sin\theta}{\cos\theta}\eps\delta.
\end{align*}
Finally, we can write the second derivatives of $\hat{\bbeta}$ :
\bg{align}
\label{r2}&\frac{1}{2}\dr^2_r\hat{\bbeta}=\frac{\sigma^3+\sigma^2\sigma''}{2S(\theta)^2}\alpha^2+\frac{C(\theta)}{2\cos\theta}\delta^2+D_0 C(\theta)-E_0S(\theta)\\
\label{rs}&\frac{1}{2}\dr^2_s\hat{\bbeta}=\frac{\sigma^3+\sigma^2\sigma''}{2S(\theta)^2}\beta^2+\frac{C(\theta)}{2\cos\theta}\eps^2+F_0 C(\theta)-3H_0S(\theta)\\
\label{s2}&\dr^2_{rs}\hat{\bbeta}=\frac{\sigma^3+\sigma^2\sigma''}{2S(\theta)^2}\alpha\beta+\frac{C(\theta)}{2\cos\theta}\delta\eps+C_0 C(\theta)-2B_0S(\theta)
\end{align}

\section{Construction of a quasimode}
In this section, we construct a quasimode and prove Theorem \ref{majgen}. More precisely, we proceed in three steps. First, we choose an appropriate scaling.
Then, we make a Fourier transform and a translation as annonced at the beginning of Section 2. We expand the new operator in powers of $h$ and look formally for a quasimode expressed as a series in powers of $h$. As we are just interested in a three terms asymptotics for the bottom of the spectrum, we will just keep the three first terms of the series. The construction of the quasimode (denoted by $\psi$) will then deeply use the results of Section 2.2. In particular, the solutions $f_0, w_0, \dr_s u_\theta$ and $\dr_s^2 u_\theta$ of equations of the form $h(\theta)v=f$  will appear in the expression of the quasimode.
Finally, we will estimate a quantity expressed as $\|(P_\A^h -\la(h))\chi\psi\|$ (where $\chi$ is a cutoff function) by using the exponential decay properties proved in Section 2.1 to control the error of the truncations and we will conclude with the spectral Theorem. 

\subsection{Rescaling}
We first make the following rescaling :
\bg{equation}
t=h^{1/2}\tilde{t}, s=h^{1/2}\tilde{s}, r=h^{1/2}\tilde{r}.
\end{equation}
We omit the tilde, we divide by $h$ and, instead of $\mj{H}^{\mj{M}}$, we consider :
\bg{align}
&H^{h,new}=n_{app}^h D_t m_{app}^h D_t+f_{11}^h(D_r+ \conj{A}_r^h)^2+f_{22}^h(D_s+\conj{A}_s^h)^2\\
\nonumber&+f_{12}^h(D_r+\conj{A}_r^h)(D_s+\conj{A}_s^h)+f_{12}^h(D_s+\conj{A}_s^h)(D_r+\conj{A}_r^h)\\
\nonumber&+h\tilde{R}_1(r,s,t)D_r+h\tilde{R}_2(r,s,t)D_s,
\end{align}
where
\bg{align*}
&\conj{A}_r^h=V_{\theta_1}+h^{1/2}P_{r,2}+hP_{r,3},\\
&\conj{A}_s^h=h^{1/2}P_{s,2}+hP_{s,3},\\
\end{align*}
and with
\bg{align*}
&n_{app}^h=1+h^{1/2}tK^M_0-h\tr(\frac{G_1^0}{2})+hR, & m_{app}^h=1-h^{1/2}tK^M_0+h\tr(\frac{G_1^0}{2})+hR\\
&f_{11}^h=1+2h^{1/2}tK_{11}-hG_{11}-hR_{11}, & f_{12}^h=2h^{1/2}K_{12}t-hG_{12}-hR_{12}\\
&f_{22}^h=1+2h^{1/2}tK_{22}-hG_{22}-hR_{22}&.
\end{align*}

\subsection{Quasimode}
We begin first by taking a partial Fourier transform in $r$, denoted by $\mj{F}$ and then we perform the translation $U_\theta$ :
$$s=\tilde{s}+\frac{\tau}{\sin\theta}.$$
to get the operator $\mj{H}^{h,new}$ which is unitarily equivalent to $H^{h,new}$ :
$$\mj{H}^{h,new}=U_\theta^{-1}\mj{F}^{-1}H^{h,new}\mj{F}U_\theta.$$
So, to get $\mj{H}^{h,new}$, we just have to replace (omitting the tilde) in $H^{h,new}$ :
$$s \mbox{ by } s+\frac{\tau}{\sin\theta} \mbox{ and } r \mbox{ by } D_\tau-\frac{1}{\sin\theta}D_s.$$
We formally expand $\mj{H}^{h,new}$ in powers of $h^{1/2}$ and we write :
$$\mj{H}^{h,new}=\sum_{j=0}^{+\infty}h^{j/2}H_j$$
with
$$H_0=D_s^2+D_t ^2+V_\theta^2$$
\bg{align*}
&H_1=(2V_\theta(\alpha t-\zeta s)+2\delta t D_s)\boxed{D_\tau}-\frac{2\zeta}{\sin\theta}V_\theta\boxed{\tau D_\tau}\\
&+(\frac{2}{\sin\theta}(\beta t-\gamma s)V_\theta+\frac{\zeta}{\sin^2\theta}(V_\theta D_s+D_s V_\theta)+\frac{2\eps}{\sin\theta}t D_s)\boxed{\tau}-\frac{\gamma}{\sin^2\theta}V_\theta\boxed{\tau^2}+\tilde{H_1},\\
\end{align*}
where $\tilde{H_1}$ is defined by :
\bg{align*}
&\tilde{H_1}=-\frac{1}{\sin\theta}(\alpha t-\zeta s)(D_s V_\theta+V_\theta D_s)+2V_\theta(\beta st-\frac{\gamma}{2}s^2)+\eta V_\theta t^2\\
&-\frac{2\delta t}{\sin\theta}D_s^2+\eps t(sD_s+D_s s)+\xi t^2 D_s+H_K
\end{align*}
and where $H_K$ is defined in (\ref{HK}). The reason why we have boxed the terms involving $\tau$ is that, after a partial scalar product in the variable $(s,t)$, we will be reduced to operators in the variable $\tau$.
The expression of $H_2$ will be discussed later.\\
We look for a quasimode expressed as a formal series in powers of $h^{1/2}$ :
$$u^h=\sum_{j=0}^{+\infty}h^{j/2}u_j$$
attached to an eigenvalue in the form :
$$\la_1^h=\sum_{j=0}^{+\infty}h^{j/2}\la_j.$$
Then, writting formally that $u^h$ satisfies 
$$\mj{H}^{h,new} u^h=\la_1^h u^h,$$
we are led to the system : 
\bg{align*}
h^0 : &\, H_0 u_0=\la_0 u_0,\\
h^{1/2} : &\, H_1 u_0+H_0 u_1=\la_1 u_0+\la_0 u_1,\\
h : &\, H_2 u_0+H_1 u_1+H_0 u_2=\la_2 u_0+\la_1 u_1+\la_0 u_2.
\end{align*}
In the next subsection, we construct a solution for this system.
\subsubsection{The coefficient of $h^0$}
$H_0$ (as an operator in the $(s,t)$ variable) has been analyzed in Section 2.
We take as $\lambda_0$ the lowest eigenvalue of $H_0$ (which was denoted before by $H(\theta)$) and we choose
\bg{equation}
u_0(\tau,s,t)=\phi_0(\tau) u_\theta(s,t)\,,
\end{equation}
with $\phi_0$ (of norm $1$) to be determined later.
\subsubsection{The coefficient of $h^{\frac 12}$}
We have to compute :
$$<H_1 u_0,u_\theta>.$$
This quantity has the structure :
$$\got{A} D_\tau \phi_0+\got{B}\tau D_\tau\phi_0+\got{C}\tau\phi_0+\got{D}\tau^2\phi_0+\got{E}\phi_0,$$
where :
\bg{align*}
&\got{A}=<(2V_\theta(\alpha t-\zeta s)+2\delta t D_s)u_\theta,u_\theta>,\\
&\got{B}=-<\frac{2\zeta}{\sin\theta}V_\theta u_\theta,u_\theta>,\\
&\got{C}=<(2(\beta t-\gamma s)V_\theta+\frac{\zeta}{\sin^2\theta}(V_\theta D_s+D_s V_\theta)+\frac{2\eps}{\sin\theta}t D_s)u_\theta,u_\theta>,\\
&\got{D}=<-\frac{\gamma}{\sin^2\theta}V_\theta u_\theta,u_\theta>,\\
&\got{E}=<\tilde{H_1}u_\theta,u_\theta>.
\end{align*}
Let us examine $\got{A}$. We recall that (\ref{f_0}) holds.
Moreover, using Lemma \ref{I1} and (\ref{zetagamma}), we find :
$$<2V_\theta(\alpha t-\zeta s)u_\theta,u_\theta>=0.$$
Thus, we have proved that $\got{A}=0$.\\
Let us now consider $\got{B}$ and $\got{D}$. By Lemma \ref{moment0}, we have : $\got{B}=\got{D}=0$.\\
Finally, let us prove that $\got{C}$ is also equal to $0$.\\
Again by (\ref{zetagamma}) and Lemma \ref{I1}, we find : 
$$<(2(\beta t-\gamma s)V_\theta u_\theta,u_\theta>=0.$$
Then, as we have just proved above, we get :
$$<\frac{2\eps}{\sin\theta}t D_s u_\theta,u_\theta>=0.$$
Finally, we recall that (\ref{ds2}) holds.
We deduce that $\la_1=\got{E}$. Let us simplify its expression.
We first observe that~:
$$<t^2 D_s u_\theta,u_\theta>=0.$$ 
Then, we get :
$$<(\alpha t-\zeta s)(D_s V_\theta+V_\theta D_s)u_\theta,u_\theta>=0.$$
Indeed, an integration by parts gives :
$$<(\alpha t-\zeta s)\dr_s (V_\theta u_\theta),u_\theta>=-<(\alpha t-\zeta s)\dr_s u_\theta,V_\theta u_\theta>,$$
where we have used Lemma \ref{moment0}. Moreover, in the same way, we get :
$$<(D_s s+sD_s)u_\theta,u_\theta>=0.$$
Consequently, we have (cf. (\ref{CBK}) and relations (\ref{an1})) :
\bg{align*}
&E=\la_1=2<V_\theta(\beta st-\frac \gamma 2)u_\theta,u_\theta>+\eta<t^2 V_\theta u_\theta,u_\theta>-\frac{2\delta}{\sin\theta}\int_{t>0} t|D_s u_\theta|^2\\
&+<H_K u_\theta, u_\theta>.
\end{align*}
So, for $u_1$, we take a solution of the equation
$$(H_0-\sigma(\theta))u_1=\la_1 u_0-H_1 u_0.$$
Gathering all the results of Section 2.2, we can write the expression of $u_1$ :
\bg{align*}
&u_1=(\frac{\alpha}{S}w_0+2i\delta f_0)\boxed{D_\tau\phi_0}+\frac{\zeta}{\sin^2\theta}\dr_s u_\theta\boxed{\tau D_\tau\phi_0}\\
&+(\frac{\beta}{S\sin\theta}w_0-i\frac{\zeta}{2\sin^3\theta}\dr_s^2u_\theta+2i\frac{\eps}{\sin\theta}f_0)\boxed{\tau\phi_0}+\frac{\gamma}{2\sin^3\theta}\dr_s u_\theta \boxed{\tau^2\phi_0}+\tilde{u}_1,
\end{align*}
where $\tilde{u}_1$ is defined by :
\beq\label{u1}
(H-\sigma)\tilde{u}_1=(\la_1-\tilde{H_1})u_\theta.
\eeq
Before analyzing the coefficient of $h$, let us state a lemma concerning $\tilde{u}_1$ :
\bg{lem}\label{Vu1}
We have the identity :
$$2\sin\theta\int_{t>0} V_\theta\tilde{u}_1 u_\theta ds dt=\int_{t>0} \dr_s(\tilde{H}_1 u_\theta) u_\theta dsdt.$$
\end{lem} 
\bg{preuve}
We take the derivative of (\ref{u1}) with respect to $s$, we multiply by $u_\theta$ and integrate.
\end{preuve}
At this step of the proof, we have not used yet the non-degeneracy condition (\ref{minimum}). In fact, this construction is enough for the proof of Theorem \ref{majgen0}.
\paragraph{The coefficient of $h$}~\\
We are interested in $\la_2$ and $u_2$ and we are led to examine the compatibility condition :
$$<H_1 u_1+H_2 u_0,u_\theta>=\la_2 \phi_0.$$
We have not given yet the explicit expression of $H_2$, but it is easy to observe that the last equation can be put into the form :
\bg{align*}
&(A_1\tau^2 D_\tau^2+A_2 \tau^4+A_3 \tau^3 D_\tau+A_4 \tau D_\tau^3\tau+A_5\tau D_\tau^2+A_6\tau^2 D_\tau+A_7\tau^3+A_8 D_\tau^3\\
&+A_9 D_\tau^2+A_{10}\tau D_\tau+A_{11}\tau^2+A_{12}D_\tau+A_{13}\tau+A_{14})\phi_0=\la_2\phi_0.
\end{align*}
Moreover, the coefficients of $K$ and $G_1$ (introduced in Section \ref{rst}) just play a priori a role in the coefficients $A_i$ for $i\in\{9,\cdots 14\}$ (for degree and scaling reasons).
In the next paragraph, let us be more accurate on that point.
\paragraph{``Reduction'' to the case of the flat metrics}~\\
We prove that the coefficients of $\tau^2, D_\tau^2$ and $\tau D_\tau$ do not depend on $G_1$ and $K$ and that we can do, in some sense, exactly as if the metrics were flat.
Before starting the analysis, let us recall that the relations (\ref{an1}) still hold so that the curvature play a role in $\xi$ and $\eta$.\\
Let us first consider the dependence of the coefficients on $K$. We begin by noticing that the one of $D_\tau^2$ has no such dependence.\\ 
Then, we consider the one of $\tau^2$. Collecting all the terms involving $K$ in $<H_1u_1,u_\theta>$, we find that they will appear only in the following coefficient (we use Lemma \ref{Vu1}) :
$$\frac{\gamma}{2\sin^3\theta}\int_{t>0} \tilde{H}_1\dr_s u_\theta u_\theta dsdt-\frac{\gamma}{2\sin^3\theta}\int_{t>0} \dr_s(\tilde{H}_1 u_\theta) u_\theta dsdt.$$
Thus, we can compute the commutator $\tilde{H}_1\dr_s-\dr_s\tilde{H}_1$ modulo the terms which do not involve the curvature ; we get :
$$\frac{\gamma}{2\sin^3\theta}(K_{11} t^2\sin\theta\cos\theta-4K_{11}t\sin\theta V_\theta-4K_{12}t\sin\theta D_s).$$
Comsequently, the coefficient is :
$$<\frac{\gamma}{2\sin^2\theta}(K_{11} t^2\cos\theta-4K_{11}tV_\theta)u_\theta,u_\theta>.$$
Then, we consider the coefficient of $\tau^2$ involving the curvature in $H_2$ :
$$\frac{\gamma K_{11}\cos\theta t^2}{2\sin^2\theta}-\frac{2\gamma K_{11}tV_\theta}{\sin^2\theta}.$$
In conclusion, the coefficient of $\tau^2$ does not depend on $K$.\\
We now analyze the coefficient of $\tau D_\tau$. Modulo terms not depending on the curvature, we see that there is only one term in $H_1$ which has to play a role in $\tau D_\tau$ and looking at the expression of the corresponding coefficient, we are reduced to the same computations as for $\tau^2.$\\
Let us now consider the dependence of the coefficients with respect to $G_1$ (see Remark \ref{notation}).
First, let us consider the term in $\tau^2$ and let us look for the terms involving $G_1$ :
$$-G_{11}^{s^2}\int_{t>0} V_\theta^2 u_\theta^2-G_{22}^{s^2}\int_{t>0}  |D_su_\theta|^2$$
and :
\beq\label{FHFb}
\frac{1}{\sin^2\theta}\left(6H\int_{t>0} s V_\theta u_\theta^2+2F\int_{t>0} t V_\theta u_\theta^2\right).
\eeq
A priori, in (\ref{FHFb}), there should be the term in $\conj{F}$, but the coefficient is essentially :
$<tD_s u_\theta, u_\theta>=0.$ 
Recalling the relations (\ref{an}), (\ref{FHFb}) gives the contribution :
$$-\frac{1}{\sin^2\theta}(G_{11}^{s^2}+G_{22}^{s^2})\sin\theta\int_{t>0} sV_\theta u_\theta^2+\frac{1}{\sin^2\theta}\cos\theta G_{11}^{s^2}\int_{t>0} t V_\theta u_\theta^2.$$
Then, we recall Remark \ref{virial2} and the coefficients of $G_{11}^{s^2}$ and $G_{22}^{s^2}$ cancel.
Secondly, we look at the terms in $D_\tau^2$ and $\tau D_\tau$ and we observe that the terms $G_{11}^{r^2}$ and $G_{22}^{r^2}$ on one hand and $G_{11}^{rs}$ and $G_{22}^{rs}$ on the other hand cancel for the same reason as for  $G_{11}^{s^2}$ and $G_{22}^{s^2}$ after having used the expressions (\ref{an2}) and (\ref{an3}).\\
Finally, the terms in $R$ (see Remark \ref{R}) playing at the scale $h$ and with degree less than $1$ with respect to $r$ and $s$ will just provide, after the Fourier transform and the translation, coefficients of $D_\tau$ and $\tau$ and, as we will see, these terms are not important to prove Theorem \ref{majgen}.\\
Now we give an explicit expression of $H_2$ (modulo terms which do not change the coefficients of $D_\tau^2$, $\tau D_\tau$ and $\tau^2$, but only those of $\tau$, $D_\tau$ and the constant)~:
\bg{align*}
&H_2=P_{r,2}\left(D_\tau-\frac{1}{\sin\theta}D_s,s+\frac{\tau}{\sin\theta},t\right)^2+P_{s,2}\left(D_\tau-\frac{1}{\sin\theta}D_s,s+\frac{\tau}{\sin\theta},t\right)^2\\
&+P_{r,3}\left(D_\tau-\frac{1}{\sin\theta}D_s,s+\frac{\tau}{\sin\theta},t\right) V_\theta+V_\theta P_{r,3}\left(D_\tau-\frac{1}{\sin\theta}D_s,s+\frac{\tau}{\sin\theta},t\right)\\
&+P_{s,3}\left(D_\tau-\frac{1}{\sin\theta}D_s,s+\frac{\tau}{\sin\theta},t\right) V_\theta+V_\theta P_{s,3}\left(D_\tau-\frac{1}{\sin\theta}D_s,s+\frac{\tau}{\sin\theta},t\right)\\
\end{align*}
where the $P_{i,j}$ are given in (\ref{polys}) (and in which we put the index $0$ at all coefficients).
The guess is that we have the following cancellations :
\beq\label{cancel}
A_i=0\quad\forall i\in\{1,\cdots,8\}.
\eeq
Then, we would be reduced to an equation of the form :
$$(A_9D_\tau^2+A_{10}\tau D_\tau+A_{11}\tau^2+A_{12}D_\tau+A_{13}\tau+A_{14})\phi_0=\la_2\phi_0.$$
Finally, writing formally that $D_\tau=r$ and $\tau=s$, we would hope to get the expression of the Hessian matrix.\\
Let us now prove the cancellations (\ref{cancel}) !
\paragraph{Coefficient of $D_\tau^3$, $\tau D_\tau^3$,}~\\
Such terms do not appear. Indeed, with the choice of gauge in (\ref{polys}), there is no $r^2$ in the homogeneous polynoms of order $2$ and no $r^3$ in the homogeneous polynoms of order $3$.
\paragraph{Coefficient of $\tau^4$}~\\
We can observe that the coefficient of $\tau^4$ in $H_2$ is :
$$\frac{\gamma^2}{4\sin^4\theta},$$ 
moreover the coefficient of $\tau^4$ in $<H_1 u_1,u_\theta>$ is :
$$-\frac{\gamma^2}{2\sin^5\theta}<V_\theta\dr_s u_\theta,u_\theta>.$$
But, taking the scalar product of (\ref{ds2}) by $u_\theta$, we obtain
$$<V_\theta\dr_s u_\theta,u_\theta>=\frac{\sin\theta}{2}.$$
\paragraph{Coefficient of $\tau^2 D^2_\tau$}~\\
The coefficient of $\tau^2 D^2_\tau$ in $H_2$ is :
$$\frac{\zeta^2}{\sin^2\theta}$$
and in $<H_1 u_1,u_\theta>$ :
$$-\frac{2\zeta^2}{\sin^3\theta}<V_\theta\dr_s u_\theta,u_\theta>.$$
\paragraph{Coefficient of $\tau^3$}~\\
The coefficient of $\tau^3$ in $H_1 u_1$ is :
\bg{align*}
&\frac{\beta\gamma}{S\sin^4\theta}<(St+Cs)\dr_s u_\theta,V_\theta u_\theta>+\frac{\gamma\zeta}{S\sin^5\theta}<(D_s V_\theta+V_\theta D_s)\dr_s u_\theta,u_\theta>\\
&+i\frac{\eps\gamma}{\sin^4\theta}<t\dr_s^2 u_\theta,u_\theta>-\frac{\gamma\beta}{S\sin^3\theta}<V_\theta w_0,u_\theta>+\frac{\gamma\zeta}{2\sin^5\theta}<V_\theta\dr_s^2 u_\theta,u_\theta>\\
&-\frac{\gamma\eps}{\sin^3\theta}<V_\theta f_0,u_\theta>.
\end{align*}
In $H_2$, the coefficient reads :
$$-\frac{\gamma\beta}{S\sin\theta}<(St+Cs)u_\theta,u_\theta>+\frac{2H}{\sin^3\theta}<V_\theta u_ \theta,u_\theta>.$$
Using Lemma \ref{Vf0}, we find that the term in $\eps\gamma$ disappear.\\
Let us consider the term in $\gamma\zeta$.
We take the derivative of (\ref{ds2}) with respect to $s$ and make the scalar product with $u_\theta$ to find : 
$$<V_\theta \dr_s^2 u_\theta,u_\theta>=0
\mbox{ and } <(D_s V_\theta,V_\theta D_s)\dr_s u_\theta,u_\theta>=0.$$
We treat the term in $\beta\gamma$ with Lemma \ref{ds2w0} and the wished cancellation follows.
\paragraph{Coefficient of $\tau^3D_\tau$}~\\
From $<H_1 u_1,u_\theta>$, we find the coefficient :
$$-\frac{\gamma\zeta}{\sin^4\theta}<V_\theta \dr_s u_\theta,u_\theta>$$
and from $<H_2 u_\theta,u_\theta>$ :
$$\frac{\zeta\gamma}{2\sin\theta}.$$
Then, we conclude exactly as for the coefficient of $\tau^2 D^2_\tau$.

\paragraph{Coefficients of $\tau D^2_\tau$ and $\tau^2 D_\tau$}~\\
Their cancellation also comes from (\ref{ds2}), Lemma \ref{Vf0} and Lemma \ref{ds2w0}.
We now study the expressions of the coefficients of the quadratic terms $\tau^2, D_\tau^2$ and $\tau D_\tau$.
\paragraph{Coefficient of $D_\tau^2$}~\\
The terms in $D_\tau^2$ coming from  $<H_1u_1,u_\theta>$ are :
\bg{align*}
&\frac{\alpha^2}{S^2}<2V_\theta(Cs+St)w_0,u_\theta>-4\delta^2<t\dr_s f_0,u_\theta>\\
&+\frac{2i\delta\alpha}{S}<2V_\theta(Cs+St)f_0,u_\theta>+\frac{2i\alpha\delta}{S}<t\dr_s w_0,u_\theta>
\end{align*}
By Lemma \ref{crossed}, the imaginary terms cancel.
Then, the terms in $D_\tau^2$ in ${<H_2u_0,u_\theta>}$ are :
\bg{align*}
&\frac{\alpha^2}{S^2}<(Cs+St)^2 u_\theta,u_\theta>+\delta^2<t^2u_\theta,u_\theta>\\
&+2D_0<tV_\theta u_\theta,u_\theta>+2E_0<sV_\theta u_\theta,u_\theta>.
\end{align*}
So, we observe, with Lemma \ref{if0} and (\ref{tVsV}), that the coefficient of $\delta^2$ is
$$-\frac{\sin\theta}{\cos\theta}\int_{t>0} ts u_\theta^2 ds dt+\int_{t>0} t^2 u_\theta^2 ds dt=\frac{1}{\cos\theta}\int_{t>0} tV_\theta u_\theta^2 dsdt=\frac{C(\theta)}{2\cos\theta}.$$
Then, with Lemma \ref{w0i} and again (\ref{tVsV}) (for the coefficients of $D_0$ and $E_0$), we have proved (see (\ref{r2})) that the coefficient of $D_\tau^2$ is :
$$\frac{1}{2}\dr^2_r\hat{\bbeta}(x_0).$$
\paragraph{Coefficient of $\tau^2$}~\\
From $<H_1 u_1,u_\theta>$ and with Lemma \ref{Vu1},we get as coefficient of $\tau^2$ :
\bg{align*}
&<\left(\frac{2}{\sin\theta}(\beta t-\gamma s)V_\theta+\frac{\zeta}{\sin^2\theta}(V_\theta D_s+D_s V_\theta)+\frac{2\eps}{\sin\theta}tD_s\right)\\
&\left(\frac{\beta}{S\sin\theta}w_0-i\frac{\zeta}{2\sin^3\theta}\dr_s^2u_\theta+2i\frac{\eps}{\sin\theta}f_0\right),u_\theta>+\frac{\gamma}{2\sin^3\theta} <[\tilde{H}_1,\dr_s]u_\theta, u_\theta>. 
\end{align*}
From $<H_2 u_0,u_\theta>$, we find the coefficient :
\bg{align*}
&\frac{1}{\sin^2\theta}<(\beta t-\gamma s)^2 u_\theta,u_\theta>-\frac{\gamma}{\sin^2\theta}<(\beta st-\frac{\gamma}{2}s^2+\frac{\eta}{2}t^2)u_\theta,u_\theta>\\
&+\frac{\zeta^2}{\sin^4\theta}<D_s^2 u_\theta,u_\theta>+\frac{\zeta}{\sin^3\theta}<(D_s(\beta t-\gamma s)+(\beta t-\gamma s)D_s)u_\theta,u_\theta>\\
&+2\frac{F_0}{\sin^2\theta}<tV_\theta u_\theta,u_\theta>+6\frac{H_0}{\sin^2\theta}<sV_\theta u_\theta,u_\theta>+2\conj{F}_0<tD_s u_\theta,u_\theta>\\
&+\frac{\eps^2}{\sin^2\theta}<t^2 u_\theta,u_\theta>+\frac{\gamma}{\sin\theta}<(\alpha t-\zeta s)D_s u_\theta,u_\theta>.
\end{align*}
We observe that $2\conj{F}_0<tD_s u_\theta,u_\theta>=0$. Then, let us gather some terms. We have first, with the same formulas as for the coefficient of $D_\tau^2$ :
$$<2(\beta t-\gamma s)V_\theta\frac{\beta}{S\sin\theta}w_0,u_\theta>+\frac{1}{\sin^2\theta}<(\beta t-\gamma s)^2 u_\theta,u_\theta>=2\frac{\sigma^3+\sigma^2\sigma'}{\sin^2\theta S(\theta)^2}\beta^2.$$
There are other terms in $\beta^2$ (or $\beta\gamma$) :
\beq\label{otherbeta2}
-\frac{\gamma}{\sin^2\theta}<(\beta st-\frac{\gamma}{2}s^2)u_\theta,u_\theta>+\frac{\gamma}{2\sin^3\theta}<[2V_\theta(\beta st-\frac{\gamma}{2}s^2),\dr_s]u_\theta,u_\theta>.
\eeq
But, we have : 
$$[2V_\theta(\beta st-\frac{\gamma}{2}s^2),\dr_s]=2\sin\theta(\beta st-\frac{\gamma}{2}s^2)-2(\beta t-\gamma s),$$
so, with (\ref{I1}), we find :
$$<[2V_\theta(\beta st-\frac{\gamma}{2}s^2),\dr_s]u_\theta,u_\theta>=<2\sin\theta(\beta st-\frac{\gamma}{2}s^2)u_\theta, u_\theta>.$$
Thus, (\ref{otherbeta2}) cancels.\\
Let us gather the terms in $\gamma\eta$ :
\beq\label{othergammaeta}
-\frac{\gamma\eta}{2\sin^2\theta}<t^2 u_\theta,u_\theta>+\frac{\gamma\eta}{2\sin^3\theta}<t^2[V_\theta,\dr_s]u_\theta,u_\theta>=0.
\eeq
As terms in $\eps^2$, we have (exactly as in the paragraph concerning $D_\tau^2$ with $\delta^2$):
$$\frac{\eps^2}{\sin^2\theta}<t^2 u_\theta,u_\theta>-\frac{4\eps^2}{\sin^2\theta}t\dr_sf_0,u_\theta>=\frac{C(\theta)}{2\sin^2\theta\cos\theta}\eps^2.$$
Let us now consider the terms in $\zeta^2$ :
\beq\label{zeta2}
\frac{\zeta^2}{\sin^4\theta}<D_s^2 u_\theta,u_\theta>+\frac{\zeta^2}{2\sin^5\theta}<(V_\theta \dr_s+\dr_s V_\theta)\dr_s^2u_\theta,u_\theta>.
\eeq
After an integration by parts, we can apply Lemma \ref{ds4} and find that (\ref{zeta2}) cancels.
Then, we have :
$$2\frac{F_0}{\sin^2\theta}<tV_\theta u_\theta,u_\theta>+3\frac{H_0}{\sin^2\theta}<sV_\theta u_\theta,u_\theta>=\frac{1}{\sin^2\theta}\left(F_0 C(\theta)-3H_0S(\theta)\right).$$
Let us now prove that all the other terms vanish.\\
So, let us first notice (Lemma \ref{crossed}) :
$$<\frac{4i\eps}{\sin^2\theta}(\beta t-\gamma s)V_\theta f_0,u_\theta>+<\frac{2i\eps\beta}{S\sin^2\theta}t\dr_sw_0,u_\theta>=0.$$
Then, we have, with Lemma \ref{Vf0} :
$$<-\frac{2\zeta\eps}{\sin^3\theta}(V_\theta \dr_s+\dr_s V_\theta) f_0,u_\theta>+<\frac{\zeta\eps}{\sin^4\theta}t\dr_s^3u_\theta,u_\theta>=0.$$
Let us write the remaining terms (for the last one, see Lemma \ref{Vu1} ) :
\bg{align*}
&<\frac{2i\zeta}{\sin^4\theta}(\beta t-\gamma s)V_\theta\dr_s^2u_\theta\,,\,u_\theta>+
<\frac{i\beta\zeta}{S\sin^3\theta}(V_\theta \dr_s+\dr_s V_\theta)w_0,u_\theta>\\
&+\frac{\zeta}{\sin^3\theta}<(D_s(\beta t-\gamma s)+(\beta t-\gamma s)D_s)u_\theta,u_\theta>\\
&+\frac{\gamma}{\sin^3\theta}<(\alpha t-\zeta s)D_s u_\theta,u_\theta>+\frac{i\gamma}{2\sin^4\theta} <[\dr_s,(\alpha t-\zeta s)(\dr_s V_\theta+V_\theta\dr_s)]u_\theta, u_\theta>.
\end{align*}
After two integrations by parts, it is easy to see that :
$$\frac{\zeta}{\sin^3\theta}<(D_s(\beta t-\gamma s)+(\beta t-\gamma s)D_s)u_\theta,u_\theta>=0.$$
Then, using Lemma \ref{ds2w0}, we get :
\bg{align*}
&-<\frac{i\zeta}{\sin^4\theta}(\beta t-\gamma s)V_\theta\dr_s^2u_\theta,u_\theta>+
<\frac{i\beta\zeta}{S\sin^3\theta}(V_\theta \dr_s+\dr_s V_\theta)w_0,u_\theta>\\
&=-i\frac{\zeta\beta}{2S\sin^3\theta}<(2C+4(Cs+St)\dr_s )u_\theta, u_\theta>=0.
\end{align*}
We have :
$$\frac{\gamma}{\sin^3\theta}<(\alpha t-\zeta s)D_s u_\theta,u_\theta>=\frac{i\zeta\gamma}{2\sin^3\theta}.$$
It is easy to see that :
\bg{align*}
&<[\dr_s,(\alpha t-\zeta s)(\dr_s V_\theta+V_\theta\dr_s)]u_\theta, u_\theta>=<(\alpha t-\zeta s)[\dr_s,(\dr_s V_\theta+V_\theta\dr_s)]u_\theta, u_\theta>\\
&=-2\sin\theta<(\alpha t-\zeta s)\dr_s u_\theta, u_\theta>=-\zeta\sin\theta.
\end{align*}
In conclusion, we have proved that the coefficient of $\tau^2$ is :
$$\frac{1}{2\sin^2\theta}\dr^2_s\hat{\bbeta}(x_0).$$
\paragraph{Coefficient of $\tau D_\tau$}~\\
From $<H_1 u_1,u_\theta>$ we get the following coefficient of $\tau D_\tau$ :
\bg{align*}
&<(2V_\theta(\alpha t-\zeta s)+2\delta tD_s) (\frac{\beta}{S\sin\theta}w_0-i\frac{\zeta}{2\sin^3\theta}\dr_s^2u_\theta+2i\frac{\eps}{\sin\theta}f_0),u_\theta>\\
&+<(\frac{2}{\sin\theta}(\beta t-\gamma s)V_\theta+\frac{\zeta}{\sin^2\theta}(V_\theta D_s+D_s V_\theta)+\frac{2\eps}{\sin\theta}t D_s)(\frac{\alpha}{S}w_0+2i\delta f_0),u_\theta>\\
&+\frac{\zeta}{\sin^2\theta}<[\tilde{H}_1,\dr_s]u_\theta,u_\theta>
\end{align*}
From $<H_2 u_0,u_\theta>$, we find the coefficient :
\bg{align*}
&-\frac{2\zeta}{\sin\theta}<(\beta st -\frac{\gamma}{2}s^2+\frac{\eta}{2})u_\theta,u_\theta>+\frac{2}{\sin^2\theta}<(\alpha t-\zeta s)D_su_\theta,u_\theta>\\
&\frac{2}{\sin\theta}<(\alpha t-\zeta s)(\beta t-\gamma s)u_\theta,u_\theta>+\frac{2\eps\delta}{\sin\theta}<t^2 u_\theta,u_\theta>\\
&+\frac{1}{\sin\theta}(4B_0\int_{t>0} sV_\theta u_\theta^2 dstdt+2C_0\int_{t>0} tV_\theta u_\theta^2 dsdt).
\end{align*}
We do not have to redo all the computations ; indeed, these are exactly the same kind of computations that we have met before. We just have to recall the relations (\ref{zetagamma}) and observe that all is divided by $\sin\theta$ and we get as coefficient :
$$\frac{1}{\sin\theta}\dr_{rs}\hat{\bbeta}(x_0).$$
In conclusion, $\phi_0$ satisfies the equation :
\bg{align}
\label{feq}&\got{S}_{\bbeta}(D_\tau,\frac{\tau}{\sin\theta})\phi_0+A_{12}D_\tau\phi_0+A_{13}\tau\phi_0+A_{14}\phi_0=\la_2\phi_0.
\end{align}
A priori $A_{12}$, $A_{13}$ and $A_{14}$ are complex numbers.
For $(c_1,c_2,d)\in\C^3$, we introduce :
\beq\label{trans}
\got{S}_{\bbeta,c_1,c_2,d}=\got{S}_\beta (D_\tau+c_1,\tau+c_2)+d.
\eeq
Then, $\got{S}_{\bbeta,c_1,c_2,d}$ is just some shifted harmonic oscillator.
After some (complex) translation to eliminate the linear terms, it is easy to put Equation (\ref{feq}) into the form :
$$\got{S}_{\bbeta,c_1,c_2,d}\tilde{\phi_0}=\la_2\tilde{\phi_0},$$
for some complex numbers $c_1$, $c_2$ and $d$.
Thus, we choose for $\tilde{\phi_0}$ the normalized eigenfunction associated to an eigenvalue $\la_2=\gamma_n(\tilde{\got{S}}_{\bbeta})+d$.
At this step, we do not know that $d$ is real. We just know that $\la_2-d$ is real.
\subsection{Upper bound}
The computations of the last section lead us to choose as a quasimode :
$$\psi=\chi(r,s,t)v^h(h^{-1/2}r,h^{-1/2}s,h^{-1/2}t),$$
where $\chi$ is a smooth cutoff function in a neighborhood of $x_0$ and where :
$$v^h=\mj{F}U_\theta u^h,$$
with 
$$u^h=u_0+h^{1/2}u_1+h u_2.$$
To get Theorem \ref{majgen0}, we would just have to keep the two first terms (which were obtained only under Assumption (\ref{minimum0})), but we do not explicit the proof which is essentially the same as the one of Theorem \ref{majgen} (see below).\\
We recall (\ref{approximation}) and we observe that (we use that $v^h$ has an exponential decay to control the commutators of $\chi$ with the derivatives by $O(h^{\infty})$) :
\bg{align}
&h^2\sum_{k,j}\|(|r|^3+|s|^3+t^3)|D_k D_j\psi|\|\leq C h^2h^{3/2}h^{-1},\\
&h\|(|r|^3+|s|^3+t^3)|\psi|\|\leq C h h^{3/2},\\
&h\sum_{j}\|(r^4+s^4+t^4)|D_j\psi|\|\leq C hh^{4/2}h^{-1/2},\\
&\|(|r|^5+|s|^5+t^5)|\psi|\|\leq C h^{5/2}.
\end{align}
Thus, we have 
$$\|\tilde{P}_\A^h\psi-\mj{H}^{\mj{M}}\psi\|\leq Ch^{5/2}.$$
Then, by construction of $u^h$ and by controlling the remainders in the expansion of $H^{h,new}$ with the exponential decay, it follows :
$$\|(\mj{H}^{\mj{M}}-(\la_0 h+\la_1 h^{3/2}+(\gamma_n(\tilde{\got{S}}_{\beta})+d) h^2))\psi\|_{L^2(m_{app}drdsdt)}=O(h^{5/2}).$$
Then, with the spectral Theorem, we have :
$$d(\sigma(P_\A^h),\la_0 h+\la_1 h^{3/2}+(\gamma_n(\tilde{\got{S}}_{\beta})+d)h^2)=O(h^{5/2}).$$
Thus, as $\sigma(P_\A^h)\subset\R$ , $\la_0\in\R$, $\la_1\in\R$, we get necessarily that  $\gamma_n(\tilde{\got{S}}_{\beta})+d\in\R.$
In particular $d$ is real and we have proved Theorem~\ref{majgen}.

\paragraph{Acknowledgments}~\\
I am very grateful to Professor Bernard Helffer for his active reading  and comments which have improved the presentation of this paper. I would also like to thank the Erwin Schr\"odinger Institute of Vienna where I have written this paper.
\bibliographystyle{alpha}
\bibliography{biblio}
\end{document}